\documentclass[preprint]{aastex}

\shorttitle{The IR SBF Hubble Constant}
\shortauthors{Jensen et al.}

\newcommand{\kms}{km\,s$^{-1}$}
\newcommand{\kmsmpc}{km\,s$^{-1}$\,Mpc$^{-1}$}
\newcommand{\Ho}{$H_{\rm 0}$}
\newcommand{\mbar}{$\overline m_{\rm F160W}$}
\newcommand{\Mbar}{$\overline M_{\rm F160W}$}
\newcommand{\vmini}{$(V{-}I)$}
\newcommand{\vminio}{$(V{-}I)_{\rm0}$}
\newcommand{\mM}{$(m{-}M)$}
\newcommand{\mMI}{$(m{-}M)_{\overline I}$}

\newcommand{\mMceph}{$(m{-}M)_{\rm Ceph}$}

\begin{document}


\title{The Infrared Surface Brightness Fluctuation 
Hubble Constant\altaffilmark{1}}
\author{Joseph B. Jensen\altaffilmark{2}}
\affil{Gemini Observatory,
	670 N. A`ohoku Pl., Hilo, HI  96720\\
	{\tt jjensen@gemini.edu}}
\author{John L. Tonry}
\affil{Institute for Astronomy, University of Hawaii\\
        2680 Woodlawn Drive, Honolulu, HI  96822\\
	{\tt jt@avidya.ifa.hawaii.edu}}
\author{Rodger I. Thompson}
\affil{Steward Observatory, University of Arizona, Tucson, AZ  85721\\
	{\tt rthompson@as.arizona.edu}}
\author{Edward A. Ajhar and Tod R. Lauer}
\affil{National Optical Astronomical Observatories\\
	P.O. Box 26732, Tucson, AZ  85726\\
	{\tt ajhar@noao.edu, lauer@noao.edu}}
\author{Marcia J. Rieke}
\affil{Steward Observatory, University of Arizona, Tucson, AZ  85721\\
	{\tt mrieke@as.arizona.edu}}
\author{Marc Postman}
\affil{Space Telescope Science Institute,
3700 San Martin Dr., Baltimore, MD  21218\\
	{\tt postman@stsci.edu}}
\and
\author{Michael C. Liu\altaffilmark{3}}
\affil{Department of Astronomy, University of California,
	Berkeley, CA  94720\\
	{\tt mliu@triscuit.berkeley.edu}}
\altaffiltext{1}{Based on observations with the NASA/ESA Hubble Space 
Telescope, obtained at the Space Telescope
Science Institute, which is operated by AURA, Inc., under NASA contract 
NAS 5-26555. }
\altaffiltext{2}{Gemini Science Fellow}
\altaffiltext{3}{Currently Beatrice Watson Parrent Fellow 
at the University of Hawaii Institute for 
Astronomy, 2680 Woodlawn Drive, Honolulu, HI  96822, 
{\tt mliu@ifa.hawaii.edu}}

\begin{abstract}
We measured infrared surface brightness fluctuation (SBF) distances 
to an isotropically-distributed sample of 16 distant galaxies with
redshifts reaching 10,000 \kms\ using the near-IR camera and 
multi-object spectrometer (NICMOS) on the {\it Hubble Space Telescope} 
(HST).  
The excellent spatial resolution, very low background, and brightness 
of the IR fluctuations yielded the most distant SBF measurements
to date.  
Twelve nearby galaxies were also observed and used to calibrate the F160W 
(1.6 \micron) SBF distance scale.  
Of these, three have Cepheid variable star distances measured
with HST and eleven have optical $I$-band SBF distance measurements.  
A distance modulus of 18.5 mag to the Large Magellanic Cloud was
adopted for this calibration.
We present the F160W SBF Hubble diagram and find a Hubble constant
\Ho$\,{=}\,76\pm 1.3$ (1-$\sigma$ statistical) $\pm 6$ (systematic) \kmsmpc.
This result is insensitive to the velocity model used 
to correct for local bulk motions.
Restricting the fit to the six most distant galaxies yields 
the smallest value of \Ho$\,{=}\,72\pm 2.3$ \kmsmpc\ consistent
with the data.  This 6\% decrease in the Hubble constant is
consistent with the hypothesis that the Local Group inhabits an
under-dense region of the universe, but is also consistent with the
best-fit value of \Ho$\,{=}\,76$ \kmsmpc\ at the 1.5-sigma level.

\end{abstract}


\section{Introduction}
The Hubble constant, \Ho, is the most fundamental of the cosmological 
parameters.  Yet in spite of its key role in our understanding of the 
universe, an accurate determination of its value eluded researchers for 
decades.  It has only been within the 
last few years that the promise of knowing \Ho\ to better than 10\% has 
been realized (Mould et al. 2000).  
The {\it Hubble Space Telescope} (HST) has occupied 
a key role in resolving the debate over the Hubble constant by 
enabling distance measurements not previously possible from the 
ground.  With HST's spatial resolution, Cepheid variable stars have been 
detected in galaxies as distant as 20 Mpc.
Cepheid distances to a variety of galaxies, including some in the
important Virgo and Fornax clusters, have provided new 
calibrations of many secondary distance indicators, including type-Ia 
supernovae (Gibson et al. 2000, Parodi et al. 2000), 
fundamental plane (Kelson et al. 2000), 
Tully-Fisher (Sakai et al. 2000), 
planetary nebulae, globular clusters, and surface brightness 
fluctuations (Ferrarese et al. 2000a).
Uniform HST Cepheid distances were collected by Ferrarese et al. (2000b).

Surface brightness fluctuations have emerged as an accurate and 
reliable distance indicator (Tonry et al. 1997, Blakeslee et al. 1999).  
HST has made it possible to not only better calibrate SBFs by 
providing Cepheid distances to calibration galaxies, but also allowed 
detection of fluctuations in half a dozen galaxies at much greater 
distances than possible from the ground 
(Lauer et al. 1998, Pahre et al. 1999, Thomson et al. 1997).  
Two additional low signal-to-noise ratio (S/N) measurements in the 
Coma cluster (Thomson et al. 1997, Jensen et al. 1999),
were the most distant SBF measurements until the current NICMOS project.

Surface brightness fluctuations have a much larger amplitude in the
near-IR than at optical wavelengths.
The Near Infrared Camera and 
Multi-object Spectrograph (NICMOS) on the HST
provides the combination of low background and high spatial resolution 
needed to measure IR SBFs beyond 100 Mpc for the first time.  The 
purpose of this study was to calibrate the F160W SBF distance scale and 
to measure \Ho\ beyond the effects of local flows.  
Reaching distances twice as large as previous SBF studies for a 
sample uniformly distributed on the sky provided immunity to many 
of the difficulties that plague all attempts to measure \Ho\ within 50 
Mpc.  

In the next section we describe the selection of the calibration
and distant galaxy samples observed.
In Section~\ref{reduxsection} we discuss the procedures used to acquire and 
reduce the data.  Section~\ref{sbfsection} describes the methods used to 
determine the SBF amplitude.  Section~\ref{calsection} discusses the 
empirical calibration of the F160W SBF distance scale and the
comparison to stellar population models.  
Section~\ref{hubblesection} presents the IR SBF Hubble diagram.
Finally, we discuss the relationship of our measurement to others which
find lower values of \Ho\ and conclude with a summary section.

\section{Sample Selection \label{samplesection}}

As part of our program to measure distances 
to redshifts of 10,000 \kms, we observed a set of nearby galaxies 
in the Leo, Virgo, and Fornax clusters.  
These observations support an empirical distance calibration
determined both using Cepheid variable star distances and the 
extensive $I$-band SBF distance survey (Tonry et al. 1997).  
The calibration galaxies cover a similar color range as the distant
galaxies used to measure \Ho\ ($I$-band SBF brightnesses show a 
systematic dependence on galaxy \vmini\ color).  
In addition to our calibration data, we discovered that several other 
NICMOS programs included F160W observations of nearby galaxies suitable 
for SBF analysis that could be used to augment our calibration.  
The most useful of these are the IR SBF survey of the Fornax cluster 
(NICMOS program 7458, J. R. Graham et al.) and the 
programs which targeted galaxies previously observed using WFPC-2 
for the purpose of measuring Cepheid distances.
We acquired raw NICMOS data from the HST archive and reduced it
using the procedures presented in this paper to guarantee a completely 
consistent calibration.  

\begin{deluxetable}{llcrrrrr}
\small
\tablecolumns{7}
\tablewidth{0pc}
\tablecaption{F160W NICMOS Observations \label{sampletable}}
\tablehead{
\multicolumn{2}{c}{Galaxy/Cluster} &
\colhead{NICMOS} &
\multicolumn{2}{c}{Galactic} &
\colhead{$A_B$} &
\colhead{$A_H$} &
\colhead{Exposure} \\
\colhead{} &
\colhead{} &
\colhead{Program} &
\colhead{long} &
\colhead{lat} &
\colhead{(mag)\tablenotemark{a}} &
\colhead{(mag)\tablenotemark{b}} &
\colhead{(sec)}
}
\startdata
\sidehead{\it Nearby Calibrators}
IC 2006  & Fornax & 7458 & 237.51 & $-$50.39 & 0.048 & 0.006 & 256 \\
NGC 1380 & Fornax & 7458 & 235.93 & $-$54.06 & 0.075 & 0.010 & 256 \\
NGC 1381 & Fornax & 7458 & 236.47 & $-$54.04 & 0.058 & 0.008 & 256 \\
NGC 1387 & Fornax & 7458 & 236.82 & $-$53.95 & 0.055 & 0.007 & 256 \\
NGC 1399 & Fornax & 7453 & 236.72 & $-$53.63 & 0.058 & 0.008 & 384 \\
NGC 1404 & Fornax & 7453 & 236.95 & $-$53.55 & 0.049 & 0.006 & 384 \\
NGC 3031 & M 81   & 7331 & 142.09 & $+$40.90 & 0.347 & 0.046 & 384 \\
NGC 3351 & Leo I  & 7330 & 233.95 & $+$56.37 & 0.120 & 0.016 & 640 \\
NGC 3379 & Leo I  & 7453 & 233.49 & $+$57.63 & 0.105 & 0.014 & 384 \\
NGC 4406 & Virgo  & 7453 & 279.08 & $+$74.63 & 0.128 & 0.017 & 384 \\
NGC 4472 & Virgo  & 7453 & 286.92 & $+$70.20 & 0.096 & 0.012 & 384 \\
NGC 4536 & Virgo  & 7331 & 292.95 & $+$64.73 & 0.079 & 0.013 & 384 \\
NGC 4636 & Virgo  & 7886 & 297.75 & $+$65.47 & 0.124 & 0.016 & 640 \\
NGC 4725 & \nodata& 7330 & 295.08 & $+$88.36 & 0.051 & 0.007 & 320\\
\sidehead{\it Intermediate-Distance Galaxies}
NGC 708  & Abell 262  & 7453 & 136.57 &$-$25.09 & 0.379 & 0.050 & 960 \\
NGC 3311 & Abell 1060 & 7820 & 269.60 &$+$26.49 & 0.343 & 0.046 & 2560 \\
IC 4296  & Abell 3565 & 7453 & 313.54 &$+$27.97 & 0.265 & 0.035 & 1920 \\
NGC 7014 & Abell 3742 & 7453 & 352.53 &$-$42.35 & 0.142 & 0.019 & 1600 \\
NGC 4709 & Centaurus  & 7453 & 302.66 &$+$21.49 & 0.512 & 0.068 & 1600 \\
NGC 5193 & (Abell 3560)\tablenotemark{c}& 7453 & 312.59 & $+$28.88 & 0.242 & 0.032 & 1920 \\
\sidehead{\it Distant Galaxies}
PGC 015524& Abell 496  & 7453 & 209.58 &$-$36.49 & 0.602 & 0.079 & 5760 \\
NGC 2832  & Abell 779  & 7453 & 191.09 &$+$44.39 & 0.073 & 0.010 & 1920 \\
IC 4051   & Abell 1656(a) & 7820 & 56.22 &$+$87.72 & 0.046 & 0.006 & 2560 \\
NGC 4874  & Abell 1656(b) & 7820 & 58.06 &$+$88.01 & 0.037 & 0.005 & 2560 \\
NGC 6166  & Abell 2199 & 7453 &  62.93 &$+$43.69 & 0.050 & 0.007 & 4160 \\
NGC 7768  & Abell 2666 & 7453 & 106.71 &$-$33.81 & 0.167 & 0.022 & 1600 \\
NGC 2235  & Abell 3389 & 7453 & 274.67 &$-$27.43 & 0.330 & 0.044 & 2048 \\
IC 4374   & Abell 3581 & 7453 & 323.14 &$+$32.85 & 0.263 & 0.035 & 1280 \\
IC 4931   & Abell 3656 & 7453 &   1.92 &$-$29.46 & 0.306 & 0.040 & 1920 \\
NGC 4073  & \nodata    & 7820 & 276.91 &$+$62.37 & 0.117 & 0.016 & 2560 \\
\enddata
\tiny
\tablenotetext{a}{SFD extinctions and assuming $A_B = 4.315E(B-V)$}
\tablenotetext{b}{Using $A_H = 0.132A_B$ (SFD)}
\tablenotetext{c}{Willmer et al. 1999 concluded that NGC 5193 is 
a foreground galaxy, and not a member of the much more
distant Abell 3560 cluster.}
\end{deluxetable}
\voffset=0in

\begin{figure}
\plotone{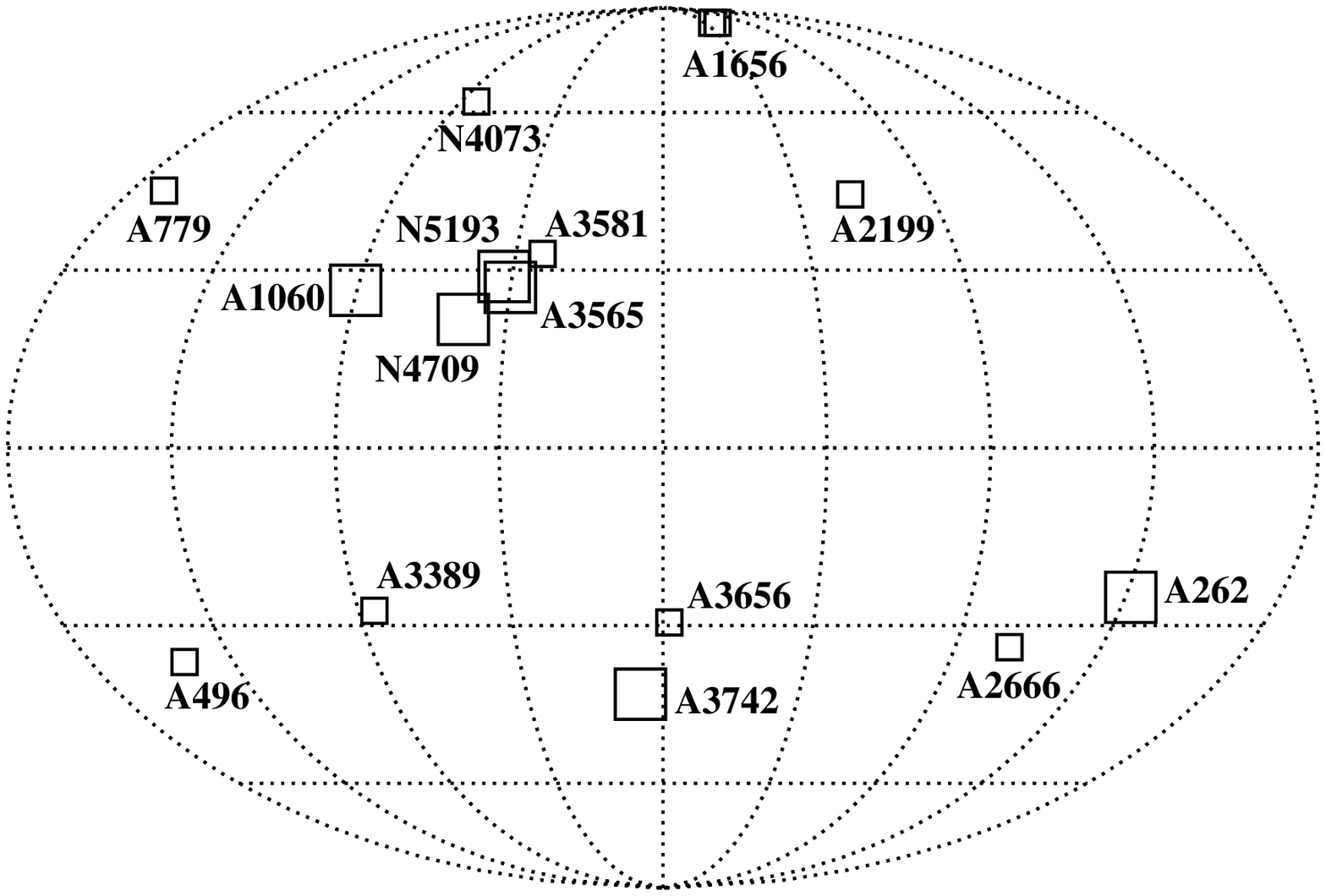}
\figcaption[]{Distribution of intermediate-distance (large squares)
and distant (small squares) galaxies of our sample, plotted
in galactic coordinates.  The distant clusters were selected to 
allow a determination of the Hubble constant that is insensitive to 
local flows.  
The intermediate-distance galaxies were chosen to provide overlap 
with optical $I$-band SBF measurements.
\label{supergal}}
\end{figure}

Because we were able to analyze 
F160W NICMOS data for galaxies with Cepheid distances,
we were not required to assume that the ellipticals and 
spirals in a given cluster are all at a common distance.
For at least a few Cepheid-bearing spirals, IR SBF measurements
are possible in their bulges.  
Using a calibration based solely on galaxies with distances measured 
both using Cepheid variables and SBFs removes the added uncertainty
in the calibration arising from the size and distribution of 
galaxies in the clusters (Tonry et al. 2000, hereafter SBF-II).

In addition to the relatively local calibrators, we also targeted five
galaxies with $I$-band SBF distances previously measured 
using WFPC-2:  the four from Lauer et al. 
(1998) and Ajhar et al. (1997), and NGC~4709 in the Centaurus cluster
(Optical SBF team, private communication).
These intermediate-distance galaxies provide overlap between 
our local calibration and the distant galaxies from which we determine \Ho.  
NGC~3311 in the Hydra cluster (NICMOS program 7820, D. Geisler et al.) 
was added to the intermediate-distance set, although it does not have an 
$I$-band SBF distance measurement.

The main focus of this study is to measure distances to 
the set of 16 galaxies (including the six intermediate-distance 
galaxies) that extend out to redshifts of 10,000 \kms.
The most distant galaxies are uniformly distributed on the sky to provide a
robust determination of \Ho\ and minimize 
sensitivity to bulk streaming motions in the local universe.  
The sample and observational data are presented in Table \ref{sampletable};
the positions of the galaxies on the sky are shown in Figure~\ref{supergal} 
in galactic coordinates.

The results presented in this paper are derived from data taken as part
of six separate NICMOS programs.  In some cases the observers in these
other programs were careful to ensure that their data would be useful
for SBF measurements.  This program demonstrates the value of the
HST archive.

\section{Observations and Data Reduction \label{reduxsection}}

All the data were acquired using the background-minimizing F160W
filter (1.6 \micron, similar to the standard $H$ filter) and the 
NIC2 camera read out in the MULTIACCUM mode.
NIC2 has a plate scale of 0.075 arcsec per pixel which gives 
a field of view 19.2 arcsec across.    
Data were reduced using modified versions of IDL 
routines developed by the NICMOS team.  During each exposure, 
the NIC2 array was read non-destructively several times, and 
intermediate images from the 
MULTIACCUM sequence were created by subtracting the initial read 
from the intermediate reads.  A dark current image from the NICMOS
team's library was then subtracted 
and pixels exhibiting non-linear response were identified.  
The differences between intermediate reads were used to identify 
pixels affected by cosmic rays, which  were 
recognizable as a change in the rate of accumulation of flux in a 
pixel and could be corrected using the unaffected sub-images.  
Remaining cosmic 
rays were fixed when the individual MULTIACCUM images were combined.  
The next step was to construct the full 
exposure by multiplying the fitted slope of flux accumulation in each 
pixel by the total exposure time, divide by the flat field, and mask bad 
pixels.  The combined MULTIACCUM images were then registered to the nearest
pixel and added together; integer-pixel registration does not
introduce correlations in the noise between
pixels that change the spatial power spectrum of the noise.
The SBF analysis assumes that the noise is uncorrelated between pixels.

Raw NICMOS images frequently have slightly different bias levels in 
each quadrant.  In the final coadded images, the background level 
mismatches between quadrants produce horizontal and vertical 
discontinuities that affect the measurement of the SBF spatial power 
spectrum.  To remove the offsets, we first processed each individual 
image and subtracted a smooth fit to the galaxy.  The differences 
between residual background levels in narrow regions on either side of the 
boundaries were measured.  Overall offsets were then computed to 
effectively add zero flux to the overall image background while 
minimizing the differences across boundaries.  Offsets were applied to 
the images prior to dividing by the flat field.  The final coadded images 
are much smoother and do not suffer from discontinuities in the 
background.

Once the relative bias levels between quadrants was removed, the overall
bias level remained uncertain.
Any such background not removed prior to dividing by the flat field
image carries the power spectrum of the flat field into the final
spatial power spectrum.  The NIC2 flat field has significant
structure, making it necessary to address the possibility that residual 
bias adds power to the measured SBF power spectrum.  
To measure the influence of residual bias levels
on the SBF measurement, we constructed an image composed of scaled
copies of that flat field added with the offsets of the dither pattern.
The resulting image was then scaled to form a ``residual bias image'' 
and added to or subtracted from the final galaxy image prior to SBF analysis.  
The SBF analysis was repeated for different
scale factors, corresponding to the likely range of residual bias values.
The most likely residual bias level was determined by trial and error:
if a residual bias correction was similar to the level of the inter-quadrant
bias adjustment, resulted in lower fluctuation amplitudes, and made the
fit to the power spectrum better over a larger range of wavenumber, then
it was adopted.  If adding a scaled residual bias image led to a worse
fit to the spatial power spectrum, or increased the fluctuation amplitude,
then no correction was adopted.  In many cases applying a residual bias
correction did not make the power spectrum fit better or worse, and no
correction was adopted. 
The influence of residual bias on the final SBF measurement was included
in the uncertainty by noting the change in the fluctuation magnitude
resulting from a range of applied residual bias levels.

Some raw NIC2 images were also affected by interference from the 
operation of the other cameras.  
Because NIC1 and NIC3 were not operated in precisely the same mode as NIC2, 
the cameras were not being reset and read at the same time.  Interference 
between cameras resulted in dark and light horizontal lines in the 
raw images that adversely affect the SBF power spectrum.  
To remove the lines, we first 
identified the affected rows in each individual image with a smooth 
galaxy profile removed.  
These rows were masked before the final image was constructed.

A few cosmic rays were energetic enough to leave a residual ghost that 
persisted for several minutes in the subsequent MULTIACCUM 
sequences.  We identified these occasional residual cosmic rays and 
masked them as well.  These, along with any cosmic rays that escaped 
detection in the MULTIACCUM sequence, were fixed using valid data for the same
location on the sky from the other images in the dither sequence.
Each individual exposure (a complete MULTIACCUM sequence) was dithered by 1.5 
arcsec, or 20 pixels.  
When the final summed images were created, we also used the 
spatial information in the dither sequence to fix the lines caused by read 
out interference.  The final images are smooth and clean, 
free from almost all the defects inherent in the raw images.

Even when great care was taken to remove cosmic rays and detector 
artifacts, one type of persistent problem proved to be difficult to 
remove from our data.  When the HST passed through the South 
Atlantic Anomaly (SAA), the NICMOS arrays were completely 
saturated with cosmic rays.  The arrays were turned off during these 
passages.  However, some of the time the arrays were restarted too 
soon after passage through the SAA, and the number of hard cosmic 
ray hits was very high.  The persistent images from these cosmic rays 
are obvious in the first MULTIACCUM sequences taken after passing 
through the SAA, and slowly decay through several subsequent 
exposures.  

\begin{figure}
\vspace{7.5in}
\includegraphics{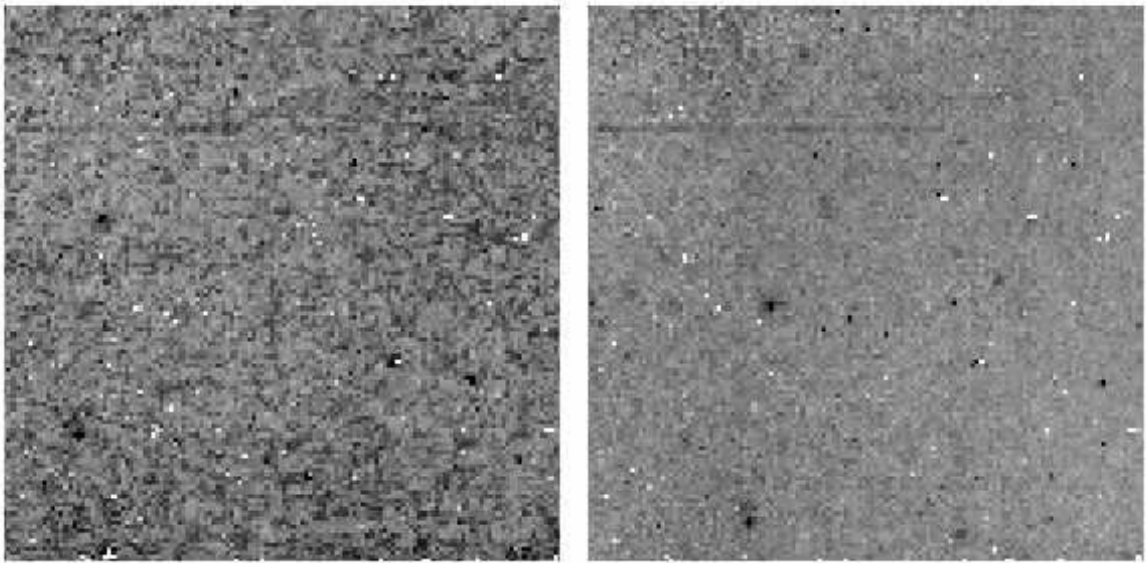}
\figcaption[]{The same quadrant (9.6 arcsec across) 
from two of the individual exposures of 
Abell 3389 are shown with the galaxy subtracted.  
The frame on the left shows a very ``wormy'' 
background in which many pixels are contaminated.  
The black speckles not seen in the image on the right
are residual images of cosmic rays.  
The right image was taken 37 minutes later,
when the residual background had largely faded.  
Both images are displayed using the {\it same} linear grayscale.
\label{worms}}
\end{figure}

The background of residual cosmic rays is best described as a 
``wormy'' pattern, with small, sometimes elongated patches of several 
pixels having significantly higher signal than surrounding regions.
An example of this wormy background is shown for one quadrant in
subsequent MULTIACCUM images in Figure \ref{worms}, in which the 
galaxy profile has been subtracted to show the background.  
The ``worms'' make up the splotchy background (left panel) and are 
distributed fairly uniformly over the array because they correspond to the 
locations of cosmic ray hits.  Worms are a serious concern for SBF 
measurements, as they are not confined to a small number of pixels.  
Their spatial power spectrum, while not exactly matching the PSF 
power spectrum, has significant power on the spatial scales used to fit 
the SBF power spectrum.  If not removed, the power in worms can 
dominate the stellar fluctuations.  The low level persistence of 
the wormy background pattern can bias the SBF measurement because 
worms add power to the fluctuation power spectrum, even when they 
are no longer obvious in the images.  

To deal with the worms, we started by examining the 
galaxy-subtracted residual images from individual MULTIACCUM sequences 
(e.g., Fig. \ref{worms}) and we excluded the badly affected images.  
The remaining question, then, is to what extent the rest of the 
images were affected.  
The power spectra from sequential exposures showed the total
power decaying to an asymptotic value, although it was difficult
to know if the contribution from worms at the asymptotic power level
was zero or not.  
Because the worms were not convolved with the PSF, it was sometimes
possible to identify wormy images from power spectra that deviated
systematically from the PSF power spectrum.  A wormy image 
has more power at high wavenumbers (small scales) and less power at 
low wavenumbers than the PSF.
To estimate the 
maximum contribution from residual worminess, we examined the 
behavior of the SBF signal as a function of the distance from the center 
of the galaxy.  The stellar SBF signal scales with the galaxy 
surface brightness, while any background fluctuation power from 
cosmic ray image persistence is uniform.  The result of background 
worminess was a fluctuation power that increased with the area of the 
region being analyzed.  We measured the SBF power in three 
or four apertures centered on the galaxy nucleus, and, assuming the stellar 
population of the galaxy is reasonably uniform (ie., the intrinsic stellar
fluctuation magnitude does not change drastically with radius), 
we applied a correction proportional to area to make the fluctuation 
measurements in the different annuli equal, if possible.  
The details of these corrections are discussed 
further in the appendix.

\section{SBF Measurements \label{sbfsection}}

\begin{figure}
\vspace{7in}
\includegraphics{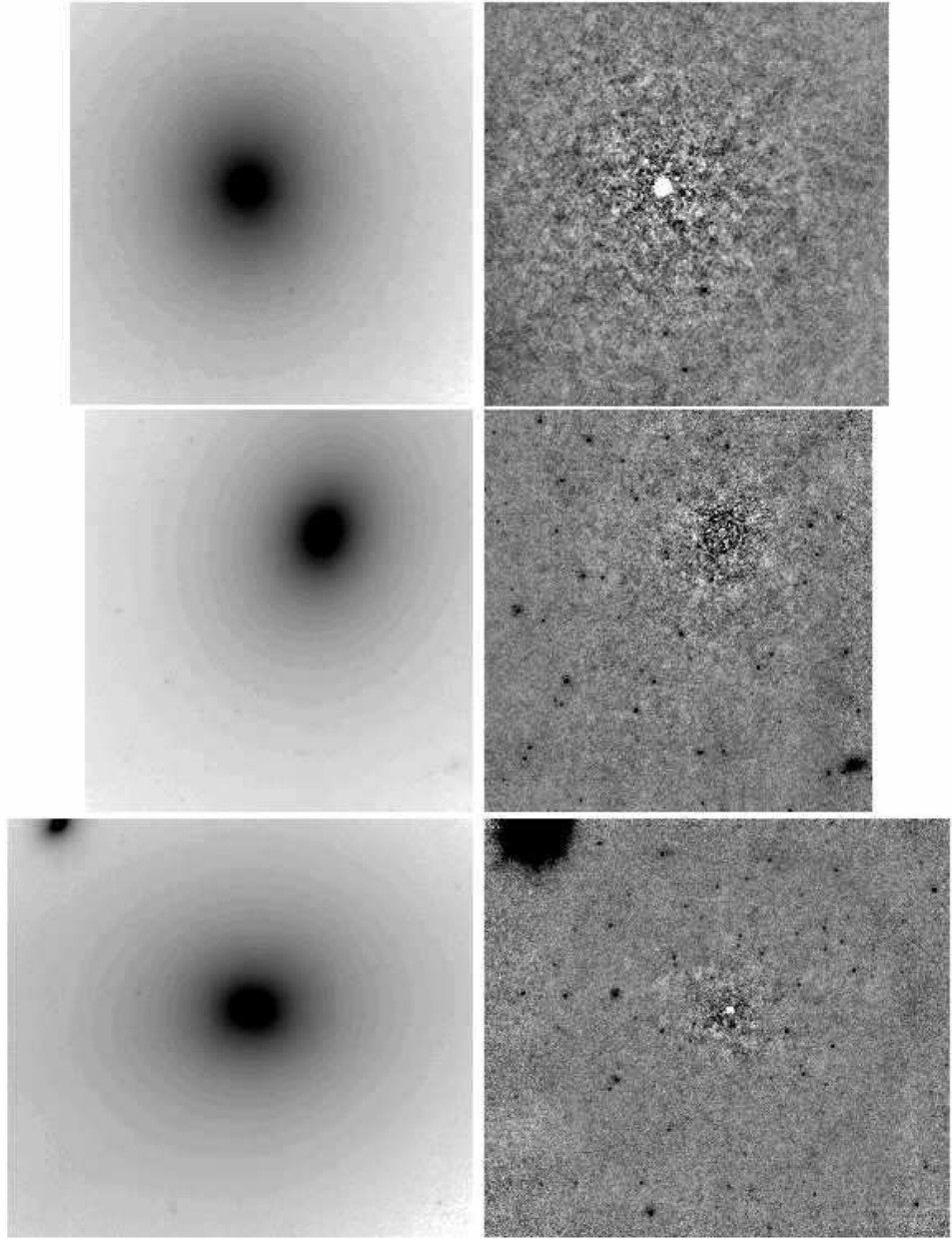}
\figcaption[]{Representative images for 
NGC~3379 (top), NGC~4709 (middle), and Abell~496 (bottom).
The images on the left are printed with a square-root stretch,
and the galaxy-subtracted images on the right are shown using
a simple linear gray scale.  Point sources in the galaxy-subtracted
images are masked prior to measuring the surface brightness 
fluctuations.  The fluctuations are easily visible as bumpiness
in the background seen in all of the field of view of NGC~3379
and near the center of Abell~496.  The globular cluster populations
can also be seen in the images of NGC~4709 and Abell~496. 
The images differ slightly in size because of the dither pattern
used, but all are approximately 20 arcmin across.  The position angle
on the sky is different for each galaxy.
\label{images}}
\end{figure}

We followed the same basic procedures for measuring SBF 
amplitudes outlined in detail in Jensen et al. (1998).  We first fitted 
and subtracted a smooth fit to the galaxy profile.  
Fluctuations can be seen in the three examples shown in Figure~\ref{images}.
Objects in the galaxy-subtracted image were identified, their 
brightnesses and 
number densities measured, and a luminosity function generated for 
globular clusters (GCs) and background galaxies (see Jensen et al. 1998
for details of the luminosity function fits).  
Objects down to the 
completeness limit were masked, and the luminosity function integrated 
beyond the completeness limit to determine the contribution to the SBFs from 
undetected globular clusters and galaxies.  
The SBF analysis is insensitive to the exact values assumed for the globular
cluster luminosity function width and peak magnitude or the galaxy 
luminosity function slope when the brightest of these populations are
well-measured and masked.
Residual large-scale 
variations in the background resulting from incomplete galaxy subtraction
were fitted and removed as well;  low 
wavenumbers ($k\,{<}\,20$) were ignored in fitting the power spectrum
because the spectrum at low wavenumbers was modified by the background
subtraction.
The fitting parameters for the galaxy profile and large-scale background
were tuned to produce the cleanest power spectrum possible.  
The uncertainties resulting from the
galaxy and background fits were determined and added in quadrature
with the other sources of uncertainty.
Dusty regions near the centers of a few of the galaxies were masked.

The SBF spatial power spectrum normalized by the mean galaxy
surface brightness was fitted with the sum of a white-noise component 
$P_1$ and the expectation power spectrum $E(k)$ scaled by the 
fluctuation power $P_0$.
$E(k)$ is a combination of the normalized PSF power spectrum and the
mask used to excise point sources and select the radial region of the
galaxy being analyzed.  The data were fitted with the function
\begin{equation}
P(k) = P_0 E(k) + P_1.
\end{equation}

The fluctuation power $P_0$ must be corrected for undetected point sources 
and residual wormy background.  These are represented by $P_r$ and
$P_g$, respectively.  The power in stellar SBFs is therefore 
\begin{equation}
P_{\rm fluc} = P_0 - P_r - P_g.
\end{equation}
$P_{\rm fluc}$ is simply a flux, and has units of electrons per pixel
per integration time.  Fluctuation powers and the relative sizes of
the $P_r$ and $P_g$ corrections are listed in Table~\ref{powertable}.
$P_{\rm fluc}$ can be transformed into an apparent fluctuation 
magnitude and corrected for galactic extinction:
\begin{equation}
\overline m = -2.5\,\log(P_{\rm fluc}) + m_1 - A_H
\end{equation}
where $m_1$ is the magnitude of a source yielding 1 e$^-$ per
total integration time on the Vega system.  
We adopted the photometric zero point
for NIC2 and the F160W filter measured by the NICMOS team of
$m_1 = 23.566 \pm 0.02$ mag, the brightness of a source which
gives 1 e$^-$s$^{-1}$.  The gain is 5.4 e$^-$ per ADU.
In this paper we have chosen to adopt the extinction values
of Schlegel et al. (1998, hereafter SFD), which are converted to the $H$-band 
extinction values assuming $A_B=4.315\,E(B{-}V)$ and $A_H=0.132\,A_B$
(SFD).
 
\begin{deluxetable}{lcrcccrcc}
\tablecolumns{9}
\tablewidth{0pc}
\tablecaption{Distant F160W SBF Measurements \label{powertable}}
\tablehead{
\colhead{Galaxy/} &
\colhead{Annulus} &
\colhead{$P_0$} &
\colhead{$P_1/P_0$} &
\colhead{$P_r/P_0$} &
\colhead{$P_g/P_0$} &
\colhead{$\xi$} &
\colhead{$P_{\rm fluc}$} &
\colhead{Notes} \\
\colhead{Cluster} & 
\colhead{(arcsec)} &
\colhead{(e$^-$pix$^{-1}$)} &
& & & & 
\colhead{(e$^-$pix$^{-1}$)} &
}
\startdata
Abell 262    & 2.4--4.8 &$ 8.8{\pm}0.3 $& 0.22 & 0.00 & 0.34 &  1.2 & \phn5.8  & d,w \\
Abell 496    & 2.4--4.8 &$11.4{\pm}0.8 $& 0.17 & 0.29 & 0.11 &  2.2 & \phn6.8  & \\
Abell 779    & 2.4--4.8 &$ 7.3{\pm}0.6 $& 0.20 & 0.28 & 0.08 &  2.2 & \phn4.6  & p \\
Abell 1060   & 2.4--4.8 &$20.7{\pm}0.5 $& 0.08 & 0.09 & 0.00 & 12.1 & 18.7 & (d) \\
Abell 1656(a)& 2.4--4.8 &$15.7{\pm}0.6 $& 0.15 & 0.41 & 0.16 &  1.4 & \phn6.8  & GC \\
Abell 1656(b)& 4.8--9.6 &$11.7{\pm}0.8 $& 0.21 & 0.15 & 0.22 &  1.5 & \phn7.4  &  \\
Abell 2199   & 2.4--4.8 &$ 9.0{\pm}0.3 $& 0.22 & 0.35 & 0.00 &  2.9 & \phn5.9  & (d) \\
Abell 2666   & 2.4--4.8 &$ 5.5{\pm}0.3 $& 0.33 & 0.00 & 0.53 &  0.6 & \phn2.6  & w \\
Abell 3389   & 2.4--4.8 &$ 8.2{\pm}0.2 $& 0.24 & 0.33 & 0.26 &  0.8 & \phn3.3  & w \\
Abell 3565   & 4.8--9.6 &$17.3{\pm}1.0 $& 0.12 & 0.06 & 0.07 &  4.7 & 15.1 & (d) \\
Abell 3581   & 2.4--4.8 &$ 8.4{\pm}0.9 $& 0.26 & 0.18 & 0.42 &  0.6 & \phn3.4  & (d),w\\
Abell 3656   & 2.4--4.8 &$ 8.0{\pm}0.6 $& 0.24 & 0.10 & 0.10 &  2.4 & \phn7.0  & p \\
Abell 3742   & 4.8--9.6 &$12.1{\pm}1.4 $& 0.22 & 0.10 & 0.05 &  3.1 & 10.3 & drift \\
NGC 4073     & 2.4--4.8 &$13.1{\pm}0.9 $& 0.15 & 0.31 & 0.25 &  1.1 & \phn5.8  & w \\
NGC 4709     & 2.4--4.8 &$20.2{\pm}0.8 $& 0.10 & 0.08 & 0.12 &  3.8 & 16.3 & w \\
NGC 5193     & 2.4--4.8 &$21.4{\pm}2.2 $& 0.09 & 0.07 & 0.00 & 10.8 & 19.9 & (d) \\
\enddata
\tablecomments{d=extensive dust, (d)=nuclear dust, w=worms, 
p=dither pattern noise.  IC~4051 (Abell 1656a) has an extensive globular
cluster population (Baum et al. 1997); Abell 3742 was affected by telescope
drift due to failure to lock onto the guide stars.}
\end{deluxetable}

Because the stellar SBF pattern is convolved with the diffraction pattern 
of the telescope and instrument, we require a good measurement of the 
point spread function (PSF) for each observation.  
If the reference PSF shape 
used does not match the PSF of the data, the fit will be poor.  If the PSF 
is not properly normalized, the photometry will not be correct.  
To ensure a good SBF measurement, we attempted to 
image a bright star concurrently with each galaxy observation 
to serve as a high-S/N PSF reference.  
The PSF star measurements were short, unguided exposures, and in some 
cases the PSF was blurred slightly by telescope drift.  Others were unusable 
because of close companions that were undetectable without the excellent 
resolution of the HST.  Still another turned out to be a compact galaxy.  
In the end, we acquired 10 good PSF measurements over the course 
of our program (spanning approximately 1 year).  
For each SBF measurement, we chose the PSF taken closest 
in time to the galaxy observation.  

The uncertainty in the SBF measurement resulting from variations in 
the PSF was determined by using all 10 PSF stars to measure the 
fluctuation magnitude for a galaxy.  The standard deviation in each 
case was added in quadrature with the other sources of uncertainty, and 
was typically between 4\% and 6\%.  The time between individual PSF
measurements was much longer than the ``breathing'' timescale of the 
telescope, so the variation between PSFs was random.  
In general, the PSF fits to the SBF data were excellent.  
While the PSF shows diffraction rings and spots that are quite different 
from the typical smooth PSF observed from the ground, the PSF
power spectrum fits the galaxy data very well, both in the tight Airy 
core (the broad, high-wavenumber component) and in the wings (the 
steeper component at low wavenumbers).  

\begin{figure}
\epsscale{0.9}
\plotone{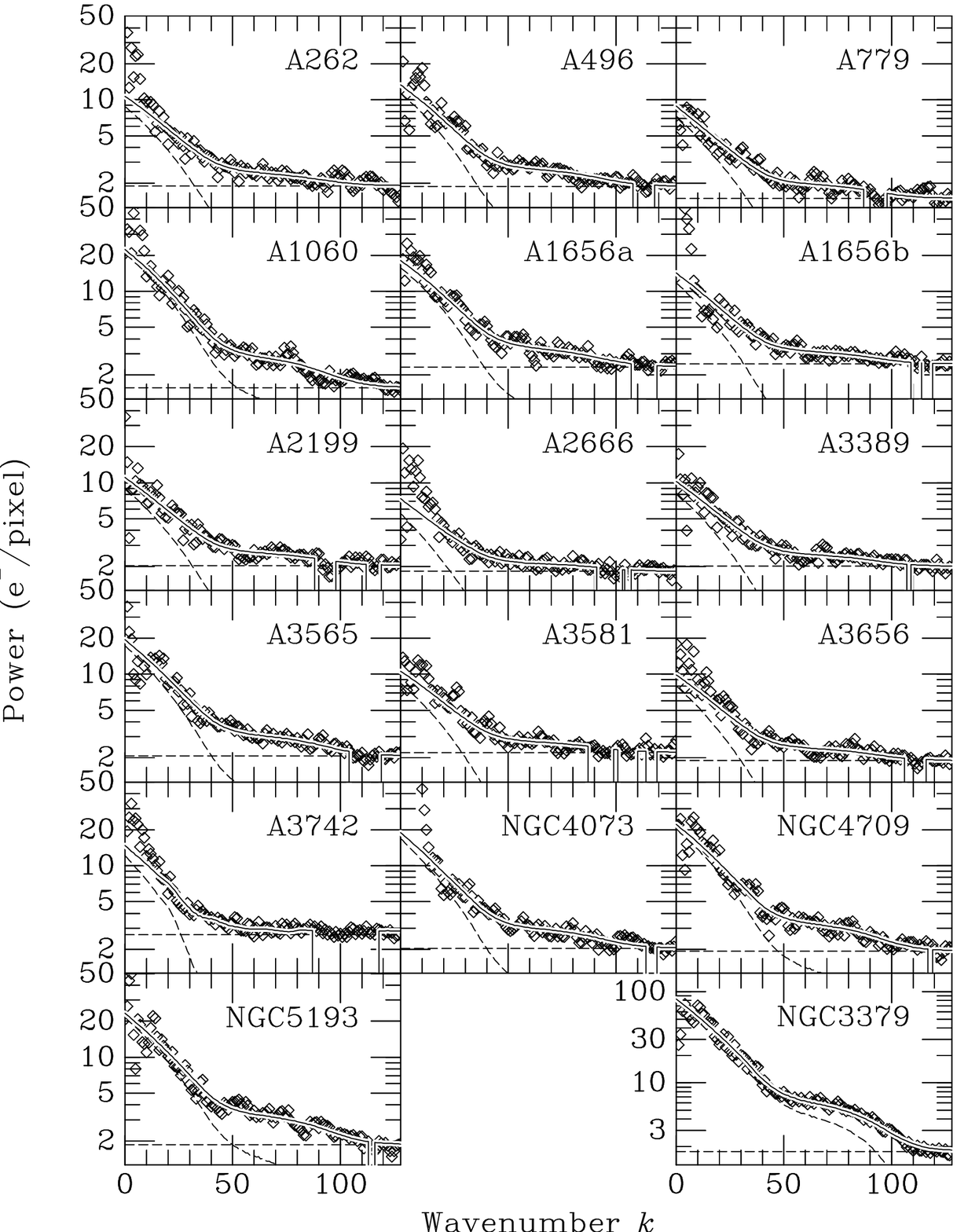}
\figcaption[]{Fluctuation spatial power spectra 
for NGC~3379 and all of the intermediate and distant galaxies.
The dashed lines indicate the white-noise component $P_1$ and the
expectation power spectrum (very nearly the power spectrum
of the normalized PSF) multiplied by the fluctuation power $P_0$.
The best fit sum is shown as a solid line.  The power spectra shown
correspond to the annular regions listed in Table~\ref{powertable}.
\label{powerfig}}
\end{figure}

We compared our snapshot PSF measurements to 16 measurements
of four stars made by the NICMOS team.  
Fluctuation measurements of our most distant galaxy
(Abell 496) were made using the library PSFs as a test case.  
The library PSF results
agreed perfectly with measurements made using our snapshot PSFs, 
and showed a somewhat smaller scatter (3.5\%). 
The smaller dispersion can be attributed to
the fact that the library PSF measurements were made on fewer stars
and while the telescope guiding was enabled.

The signal-to-noise ratio of an SBF measurement is best quantified as 
\begin{equation}
\xi = P_{\rm fluc} / (P_1 + P_g).
\end{equation}
Jensen et al. (1998) showed that $\xi$ is a good figure of merit for
IR SBF measurements.  Values of $\xi$ less than unity 
indicate measurements that are unreliable.  The higher $\xi$, the
better the SBF measurement.  
Galaxies with $\xi{<}1$ are
necessarily those for which the correction for globular clusters ($P_r$) 
or residual cosmic rays ($P_g$) are large.  
$P_0/P_1$, while sometimes used as a measure of SBF S/N, significantly
overestimates the true S/N because $P_0$ contains the contributions from
these other sources of variance.  

In several cases, the relative contributions of stellar SBFs 
($P_{\rm fluc}$), background worminess ($P_g$), and globular 
clusters ($P_r$) to the power spectrum were difficult to untangle.
Table~\ref{powertable} lists the powers measured (in electrons per
total integration time) for each galaxy and the relative levels
of the $P_1$, $P_r$, and $P_g$ contributions.  The fluctuation S/N ratio 
($\xi$) is listed for each galaxy, and the power spectrum for each
annulus listed in Table~\ref{powertable} is plotted in 
Figure~\ref{powerfig}.  Fluctuation measurements were made
in three annuli for each galaxy, and the results compared.  
The inner annulus spanned a radial region from 1.2 to 2.4 arcsec, 
the middle annulus from 2.4 to 4.8 arcsec, and the outer annulus from 4.8 to
9.6 arcsec.
In the appendix we discuss the SBF measurements for each 
intermediate and distant galaxy individually.

\section{Calibration of the F160W SBF Distance Scale \label{calsection}}

\subsection{Absolute Fluctuation Magnitudes}

Apparent fluctuation magnitudes for the nearby calibrator galaxies
were combined with previously measured distance moduli to empirically
determine the absolute brightness of F160W fluctuations \Mbar.
Most of the calibration galaxies are giant ellipticals, and we adopt the
distances from the $I$-band SBF survey for them (Optical SBF team,
private communication).  The $I$-band SBF
distances were calibrated using Cepheid distances to a handful of
spiral galaxies for which $I$-band SBF analysis was possible in the bulges
(SBF-II).  Thus the $I$-band SBF distances used to calibrate
the F160W distance scale are based on SBF and Cepheid distances
to individual galaxies, and do not assume common distances for different
galaxies within a cluster or group.  Optical SBF distance moduli and F160W
fluctuation magnitudes are listed in Table~\ref{caltable}.

\begin{deluxetable}{lcccccc}
\tablecolumns{7}
\tablewidth{0pc}
\tablecaption{F160W SBF Calibration Measurements \label{caltable}}
\tablehead{
&&&
\multicolumn{2}{c}{$I$-band SBF Distances} &
\multicolumn{2}{c}{Cepheid Distances}
\\
\colhead{Galaxy/} &
\colhead{\mbar} &
\colhead{\vminio} &
\colhead{\mMI} &
\colhead{\Mbar} &
\colhead{\mMceph\tablenotemark{a}} &
\colhead{\Mbar} \\
\colhead{Cluster} &
\colhead{(mag)} &
\colhead{(mag)} &
\colhead{(mag)} &
\colhead{(mag)} &
\colhead{(mag)} &
\colhead{(mag)} 
}
\startdata
\sidehead{\it Nearby Calibrators}
IC 2006  & $26.58\pm0.05$ & $1.183\pm0.018$ & $31.59\pm0.29$ & $-5.01\pm0.29$ & \nodata & \nodata \\
NGC 1380 & $26.40\pm0.05$ & $1.197\pm0.019$ & $31.32\pm0.18$ & $-4.92\pm0.18$ & \nodata & \nodata \\
NGC 1381 & $26.52\pm0.10$ & $1.189\pm0.018$ & $31.28\pm0.21$ & $-4.76\pm0.23$ & \nodata & \nodata \\
NGC 1387\tablenotemark{b} & $26.0\phn\pm0.7\phn$ & $1.208\pm0.047$ & $31.54\pm0.26$ & $-5.6\phn\pm0.8\phn$ & \nodata & \nodata \\
NGC 1399 & $26.76\pm0.04$ & $1.227\pm0.016$ & $31.50\pm0.16$ & $-4.74\pm0.15$ & \nodata & \nodata \\
NGC 1404 & $26.66\pm0.08$ & $1.224\pm0.016$ & $31.61\pm0.19$ & $-4.95\pm0.19$ & \nodata & \nodata \\
NGC 3031 & $22.96\pm0.05$ & $1.187\pm0.011$ & $27.96\pm0.26$ & $-5.00\pm0.26$ & $27.80\pm0.08$ & $-4.84\pm0.09$ \\
NGC 3351\tablenotemark{c} & $25.16\pm0.07$ & $1.225\pm0.014$ & \nodata & \nodata & $30.01\pm0.08$ & $-4.85\pm0.10$ \\
NGC 3379 & $25.23\pm0.08$ & $1.193\pm0.015$ & $30.12\pm0.11$ & $-4.89\pm0.13$ & \nodata & \nodata \\
NGC 4406 & $26.23\pm0.06$ & $1.167\pm0.008$ & $31.17\pm0.14$ & $-4.94\pm0.14$ & \nodata & \nodata \\
NGC 4472 & $26.23\pm0.04$ & $1.218\pm0.011$ & $31.06\pm0.10$ & $-4.83\pm0.09$ & \nodata & \nodata \\
NGC 4536\tablenotemark{b,c} & $25.43\pm0.12$ & $1.20\phn\pm0.07\phn$ & \nodata & \nodata & $30.95\pm0.07$ & $-5.52\pm0.14$ \\
NGC 4636 & $26.07\pm0.08$ & $1.233\pm0.012$ & $30.83\pm0.13$ & $-4.76\pm0.15$ & \nodata & \nodata \\
NGC 4725 & $25.69\pm0.10$ & $1.209\pm0.023$ & $30.61\pm0.34$ & $-4.92\pm0.35$ & $30.57\pm0.08$ & $-4.88\pm0.12$ \\

\sidehead{\it Intermediate-Distance Galaxies\tablenotemark{d}}
Abell 262\tablenotemark{b} & $29.06$ & $1.275\pm0.015$ & $33.99\pm0.20$ & $-4.96$ & \nodata & \nodata \\
Abell 3565 & $28.79$ & $1.199\pm0.015$ & $33.69\pm0.16$ & $-4.92$ & \nodata & \nodata\\
Abell 3742 & $29.03$ & $1.248\pm0.015$ & $34.00\pm0.15$ & $-5.00$ & \nodata & \nodata \\
NGC 4709 & $28.48$ & $1.221\pm0.015$ & $33.04\pm0.17$ & $-4.58$ & \nodata & \nodata \\
NGC 5193 & $28.49$ & $1.208\pm0.015$ & $33.51\pm0.15$ & $-5.04$ & \nodata & \nodata \\
\enddata
\tablenotetext{a}{Ferrarese et al. 2000a}
\tablenotetext{b}{Significant dust; exluded from \Mbar\ calibration.}
\tablenotetext{c}{\vmini\ color measured in a region matching the NICMOS 
field of view; the others were measured in larger apertures.}
\tablenotetext{d}{F160W SBF magnitudes, distances, and  uncertainties 
for the intermediate set are listed in Table~\ref{distancetable}. 
$I$-band SBF distance moduli from Lauer et al. 1998 have been
recalcuated using SFD extinction corrections. 
Absolute F160W fluctuation magnitudes are listed here for comparison to 
the calibration sample.}
\end{deluxetable}

Four Cepheid-bearing galaxies were observed as part of other NICMOS programs, 
which allowed us to bypass the $I$-band SBF calibration altogether.  
We determined reliable \Mbar\ values for three of the galaxies using HST 
Cepheid distances;  the fluctuation measurement in NGC~4536 was contaminated
by clumpy dust, and it was excluded from the calibration.  
Cepheid distances are also compiled in Table~\ref{caltable}.

The measured apparent fluctuation magnitudes are very robust. 
Because slewing between targets and acquiring guide stars takes a 
significant fraction of an HST orbit, only one or two calibration 
galaxies could be observed in one orbit.
Each integration was at least 256 s, far longer than the minimum time 
needed to measure SBFs with NICMOS at distances less than 20 Mpc.  
As a result, the fluctuations were very strong in the calibration images, 
and the corrections for undetected globular clusters and worms were 
insignificant.  
The S/N ratios of the calibration measurements were $\xi{=}15$ to 50. 

\begin{figure}
\plotone{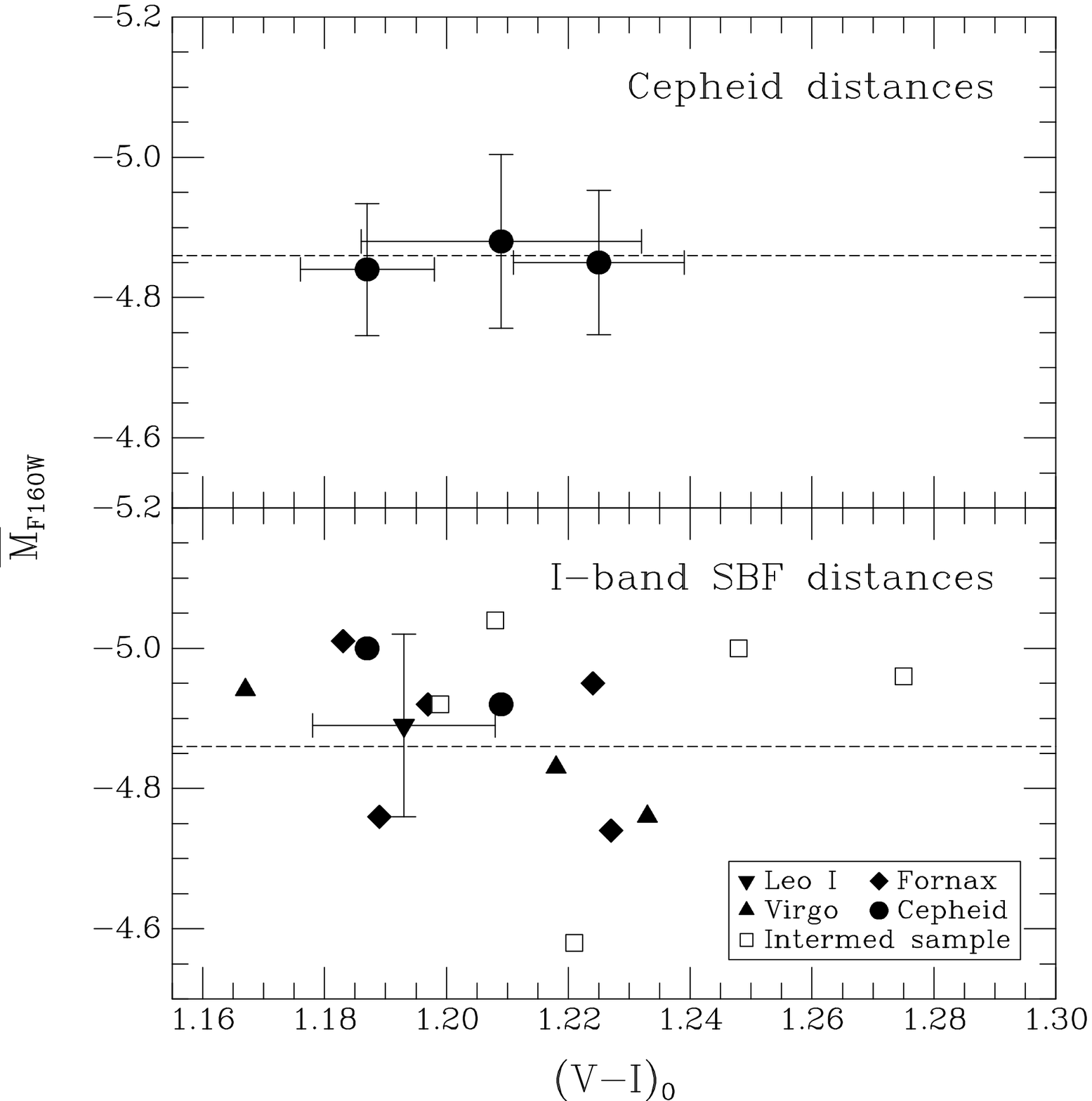}
\figcaption[]{Absolute fluctuation magnitudes \Mbar\ 
as a function of galaxy color \vminio, corrected for extinction.
The upper panel shows the three galaxies for which we have reliable
F160W fluctuation magnitudes and Cepheid distances.  In the lower
panel we plot \Mbar\ for the calibrators using their $I$-band SBF distances.
Error bars are shown for NGC~3379, and are typical for the set of
calibrators.  
Note that two of the galaxies with Cepheid distances also have $I$-band
SBF measurements. 
The horizontal line indicates the calibration \Mbar$\,{=}\,-4.86\,{\pm}\,0.05$
adopted here.  The open symbols in the lower panel are the
intermediate-distance galaxies with $I$-band SBF distances measured
using WFPC-2 and have similar uncertainties.  
They were not used in generating the calibration fit
(including them only changes the fitted \Mbar\ by 0.015 mag).  
They are shown here to demonstrate that the calibration derived using 
relatively nearby spirals and ellipticals applies equally well to the 
brightest cluster galaxies of our distant set.
\label{calibration}}
\end{figure}

Absolute fluctuation magnitudes are plotted
as a function of \vmini\ color in Figure~\ref{calibration}.
NGC~1387 and NGC~4536 are excessively dusty, and the extra spatial
power from the clumpy dust leads to the anomalously bright fluctuation
magnitudes measured (they lie outside the range plotted in 
Figure~\ref{calibration}).  The dust is easily seen in the NICMOS images,
and these two galaxies were rejected from further consideration based
on morphology, not their bright fluctuation magnitudes.  
They were excluded from the calibration fits.  
The top panel of Figure~\ref{calibration} shows \Mbar\ derived using
Cepheid distances, and the lower panel shows those calculated using
$I$-band SBF distances.

The best fits for the larger $I$-band SBF calibration and the direct
Cepheid calibration are practically identical.
A weighted fit (including the uncertainties both in \Mbar\ and \vmini)
has no significant slope in \Mbar\ for galaxies redder than
\vmini${>}1.16$.  
We therefore adopt a uniform \Mbar\ calibration for galaxies in this
color range, and note that none of the distant galaxies are likely to
have colors bluer than \vmini${=}1.16$ (as described in the next
section).  The $I$-band SBF distances give
\begin{equation}
\overline M_{\rm F160W} = -4.86\pm0.05\,{\rm mag}
\end{equation}
with an rms scatter of 0.08 mag. 
The calibration derived using only the three Cepheid measurements
is indistinguishable ($-4.85\pm0.06$ mag).
The small scatter in values of \Mbar\ for this color range
is remarkable, and emphasizes the potential IR SBFs have as a precision
distance indicator and probe of stellar populations.

Five additional galaxies from the intermediate-distance sample are
plotted in Figure~\ref{calibration} with open symbols.  
These galaxies have $I$-band SBF distances measured using HST.
They were not included in the calibration fit; instead, their
F160W SBF distances were derived using the calibration and they 
were included in the computation of \Ho.
If they had been used as calibrators, the calibration would have been 
0.015 mag brighter, which is entirely consistent given the standard
deviation of 0.05 mag observed in the \Mbar\ fit.
The intermediate set is presented in Figure~\ref{calibration} to 
demonstrate the overlap between our calibration sample and the distant 
galaxies from which \Ho\ is derived.
While both intermediate-distance galaxies redder than \vminio$\,{=}\,1.24$
have brighter than average absolute fluctuation magnitudes, we
cannot assume that redder galaxies have intrinsically brighter
fluctuations.  In fact, the results of Jensen et al. (2000) suggest
that \Mbar\ gets fainter with increasing \vmini.  
Different stellar population models (Sec. \ref{modelsection}) provide
contradictory predictions for fluctuation magnitudes in galaxies
redder than \vmini$\,{=}\,1.24$.  
At this point, we take the conservative approach and adopt a uniform
calibration for all the distant galaxies, relying on the overlap
(albeit with significant scatter) between the calibrators and the
intermediate set.
The consistency of \Mbar\ values shown in   
Figure~\ref{calibration} suggests that there are probably no significant
stellar population differences between the distant brightest cluster
galaxies, the nearby ellipticals, and the bulges of the Cepheid-bearing
spirals that produce large variations in the F160W absolute
fluctuation magnitudes.

The two versions of the calibration presented here are not independent;
both rely on many of the same Cepheid calibrators and are subject
to the same systematic uncertainties of the Cepheid distance scale.  
These significant uncertainties are very much the topic of current
debate, and include the issues of the distance to the Large Magellanic
Cloud (Mould et al. 2000), 
metallicity corrections to the Cepheid distance scale
(Kennicutt et al. 1998; Ferrarese et al. 2000a),
and blending of images in the most distant Cepheid measurements
(Ferrarese et al. 1998, 2000c; Gibson, Maloney, \& Sakai 2000; 
Stanek \& Udalski 2000; Mochejska et al. 2000).
This study adopts a distance modulus to the LMC of 18.50 mag. 
The Cepheid distances adopted are those of Ferrarese et al. (2000b),
without the metallicity correction described in Kennicutt et al. (2000).

\subsection{\vmini\ Colors} 

As in the optical $I$-band, \Mbar\ shows a dependence on 
\vmini\ color such that bluer ellipticals have intrinsically brighter 
fluctuations.  Stellar population models predict a breaking of the age 
and metallicity degeneracy in the near-IR, 
and the observed slope of \Mbar\ with \vmini\ 
reveals differences between old, metal-poor populations and young, 
metal-rich galaxies.  A sample of NICMOS SBF measurements in galaxies
spanning a wide range in \vmini\ is presented in a companion paper
(Jensen et al. 2000) in which stellar population issues are explored.
The slope in \Mbar\ with color among the redder ellipticals 
\vmini${>}1.16$ is insignificant (Fig.~\ref{calibration}).  
\vmini\ colors have been measured for 7 of the 
16 distant galaxies in our sample (Lauer et al. 1998), 
and all are significantly redder than \vmini$\,{=}\,1.16$.  
Estimates of the \vmini\ colors for the rest of the distant sample 
were made by finding the best-fitting relationship between \vmini\
and ($B{-}R$) for the 7 galaxies for which both colors are known,
and then applying the relationship to the ($B{-}R$) data taken from 
Lauer \& Postman (1995).
All of the estimated \vmini\ colors are significantly redder than 
1.16 as well.
The mean estimated \vmini\ is 1.246 mag with a standard deviation of 0.027 
(averaging all 14 galaxies with known ($B{-}R$) colors);
the mean for the 7 galaxies with measured \vmini\ colors is
1.255 mag.  
Measured or estimated \vminio\ colors listed in Table \ref{distancetable} 
have been corrected for extinction and redshift.  
We therefore feel secure adopting the calibration determined for galaxies 
redder than 1.16 for the distant sample.  
Distances to bluer galaxies will require a reliable \vminio\ 
color measurement and the full color--\Mbar\ 
relation to be presented in Jensen et al. (2000). 

\begin{deluxetable}{lccccccccr}
\small
\tablecolumns{10}
\tablewidth{0pc}
\tablecaption{F160W SBF Distances and Velocities\label{distancetable}}
\tablehead{
\colhead{Galaxy/} &
\colhead{\mbar} &
\colhead{max} &
\colhead{min} &
\colhead{\vminio\tablenotemark{a}} &
\colhead{$k(z)$\tablenotemark{b}} &
\colhead{\mM\tablenotemark{c}} &
\colhead{$d$} &
\colhead{$v_{\rm CMB}$\tablenotemark{d}} &
\colhead{$N$\tablenotemark{e}} \\
\colhead{Cluster} &
\colhead{(mag)} &
\colhead{(mag)} &
\colhead{(mag)} &
\colhead{(mag)} &
\colhead{(mag)} &
\colhead{(mag)} &
\colhead{(Mpc)} &
\colhead{(\kms)} & 
}
\startdata
A262    & $29.06{\pm}0.08$ & $+0.06$ & $-0.36$ & $1.275{\pm}0.015$ & 0.027 & $33.89{\pm}0.10$ & \phn60  & 4618 & 128 \\
A496    & $30.80{\pm}0.09$ & $+0.13$ & $-0.18$ & (1.21)            & 0.052 & $35.61{\pm}0.11$ & 132 & 9799 & 147 \\
A779    & $30.10{\pm}0.09$ & $+0.00$ & $-0.14$ & (1.28)            & 0.036 & $34.92{\pm}0.11$ & \phn97  & 7089 & 59 \\
A1060   & $28.86{\pm}0.07$ & $+0.26$ & $-0.01$ & (1.28)            & 0.020 & $33.70{\pm}0.08$ & \phn55  & 4061 & 102 \\
A1656(a)& $30.00{\pm}0.12$ & $+0.16$ & $-0.34$ & $1.297{\pm}0.037$ & 0.039 & $34.82{\pm}0.13$ & \phn92  & 7245 & 377 \\
A1656(b)& $29.91{\pm}0.08$ & $+0.00$ & $-0.32$ & $1.295{\pm}0.037$ & 0.039 & $34.73{\pm}0.10$ & \phn88  & 7244 & 377 \\
A2199   & $30.68{\pm}0.11$ & $+0.00$ & $-0.64$ & \nodata           & 0.049 & $35.49{\pm}0.12$ & 125 & 8935 & 121 \\
A2666   & $30.50{\pm}0.12$ & $+0.69$ & $-0.20$ & (1.25)            & 0.044 & $35.32{\pm}0.13$ & 116 & 7888 & 30 \\
A3389   & $30.49{\pm}0.10$ & $+1.14$ & $-0.54$ & (1.24)            & 0.042 & $35.31{\pm}0.12$ & 115 & 8105 & 39 \\
A3565   & $28.79{\pm}0.08$ & $+0.07$ & $-0.02$ & $1.199{\pm}0.015$ & 0.019 & $33.63{\pm}0.09$ & \phn53  & 4142 & 15 \\
A3581   & $29.97{\pm}0.08$ & $+0.24$ & $-0.77$ & (1.27)            & 0.035 & $34.80{\pm}0.09$ & \phn91  & 6778 & 29 \\
A3656   & $29.62{\pm}0.07$ & $+0.13$ & $-0.07$ & (1.24)            & 0.031 & $34.45{\pm}0.09$ & \phn78  & 5607 & 18 \\
A3742   & $29.03{\pm}0.11$ & $+0.05$ & $-0.16$ & $1.248{\pm}0.015$ & 0.026 & $33.86{\pm}0.12$ & \phn59  & 4801 & 20 \\
N4073   & $30.16{\pm}0.12$ & $+0.36$ & $-0.67$ & \nodata           & 0.032 & $34.99{\pm}0.13$ & \phn99  & 6306 & 1 \\
N4709   & $28.48{\pm}0.07$ & $+0.00$ & $-0.14$ & $1.221{\pm}0.015$ & 0.017 & $33.32{\pm}0.08$ & \phn46  & 4905 & 1 \\
N5193   & $28.49{\pm}0.06$ & $+0.36$ & $-0.15$ & $1.208{\pm}0.015$ & 0.019 & $33.33{\pm}0.08$ & \phn46  & 3920 & 1 \\
\enddata

\tablenotetext{a}{\vminio\ colors were derived from WFPC-2 data 
(this study and Lauer et al. 1998) and have had SFD extinction and redshift $k$-corrections 
applied.  Values listed in parentheses are estimates derived from 
ground-based photometry (Postman \& Lauer 1995) in much larger apertures 
(see text).}
\tablenotetext{b}{Liu et al. 2000 model values}
\tablenotetext{c}{Uncertainties include only the Gaussian uncertainties
combined in quadrature with the calibration uncertainty.} 
\tablenotetext{d}{Heliocentric velocities from Postman \& Lauer 1995 
(with additional measurements included) were transformed into the 
cosmic microwave background frame as described in Lineweaver et al. 1996.}
\tablenotetext{e}{Number of galaxies used to determine the cluster velocity.}
\end{deluxetable}

\subsection{Comparison with Single-Burst Stellar Population Models
\label{modelsection}}

If the intrinsic luminosity of the brightest stars in a population is
known, fluctuation distances can be determined directly without
an empirical calibration based on another distance indicator.
Stellar population models can be used to compute theoretical
absolute fluctuation magnitudes by determining the second moment of 
the luminosity function for an ensemble of stars of a particular age 
and metallicity.  
In practice, we adopt the empirical calibration because of the 
uncertainties involved in modeling populations and because of the
variations in the ages and metallicities of real galaxies.  
Nevertheless, it is useful to compare the empirical calibration 
to the theoretical predictions of stellar population models.  

\begin{figure}
\plotone{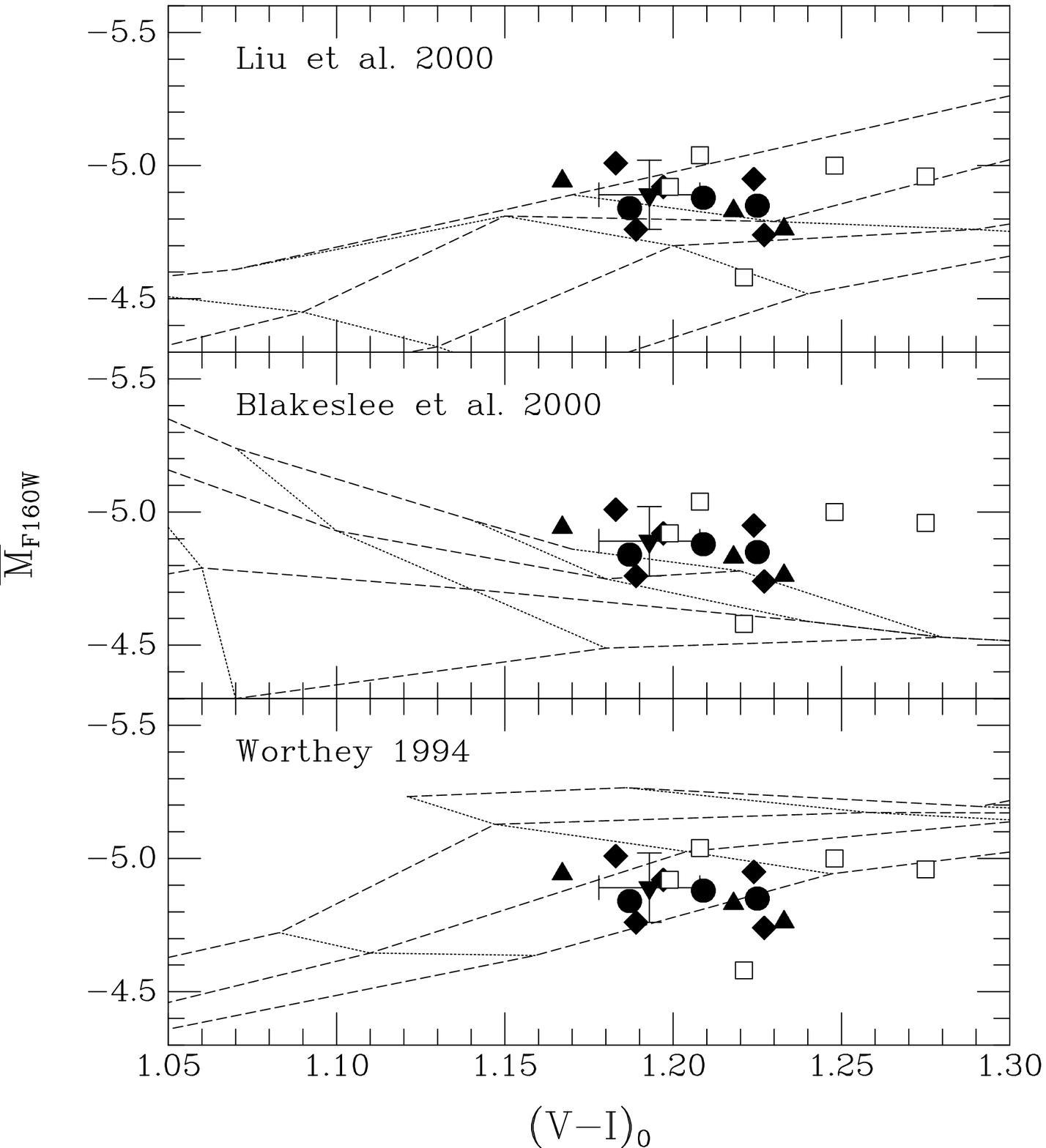}
\figcaption[]{The same data shown in Figure \ref{calibration}
plotted over three different sets of stellar population
models, as described in the text.  The distances used
are $I$-band SBF distances except for the three galaxies with
Cepheid distances.  The symbol definition is the same as in 
Figure~\ref{calibration}.  The bottom dashed lines
are 17 Gyr isochrones, followed upward by the 12, 8, and 5 Gyr (top) 
models (sloping up and to the right in the Liu et al. and Worthey models).
The data points are clustered around the solar metallicity tracks
in the Liu et al. models, and the next dotted
line down and to the left is the [Fe/H]$\,{=}\,{-}0.4$ set of models.
In the middle panel, the points are closest to the 
[Fe/H]$\,{=}\,0.2$ models, and extend down to the solar metallicity line.
The data points lie closest to the Worthey models with metallicity
[Fe/H]$\,{=}\,{-}0.25$; the lower left line indicates the 
[Fe/H]$\,{=}\,{-}0.5$ models.
\label{models}}
\end{figure}

In Figure \ref{models} we plotted the same calibration data points 
shown in Figure \ref{calibration} over three different sets of models.  
The top panel shows the recent model predictions
of Liu, Charlot, \& Graham (2000) for the F160W filter.  
These are the same models used to compute the redshift corrections
$k(z)$ to our fluctuation magnitudes.  The models plotted in the middle
panel were taken from Blakeslee, Vazdekis, \& Ajhar (2000).
These models were computed for the $H$-band filter and shifted
to F160W using the relation:
\begin{equation}
\overline M_{\rm F160W} = \overline M_H +0.1(\overline M_J - \overline M_K)
\end{equation}
(J. Blakeslee, private communication, and Stephens et al. 2000).
Finally, Worthey's (1994) models are plotted in the bottom panel
for the F160W filter (G. Worthey, private communication).
In all three sets of models, dashed lines indicate isochrones ranging
from approximately 5 Gyr at the top to 17 Gyr on the bottom.  The 
fine dotted lines indicate models of constant metallicity.
In the Liu et al. case, the SBF measurements straddle the
solar-metallicity line, and the next line to the left is 
[Fe/H]\,${=}\,{-}0.4$.  In the center panel, the points are closest
to the [Fe/H]\,${=}\,0.2$ models of Blakeslee et al. (2000) and reach
down to the solar metallicity line.  The transformation to the F160W
scale, and hence the vertical position of the models, is somewhat 
uncertain however.
In the bottom panel, the points are closest to the [Fe/H]\,${=}\,{-}0.25$
line; the next metallicity line down is [Fe/H]\,${=}\,{-}0.5$.
The calibration data presented here cover a limited range in color, 
and appear consistent with stellar populations near solar metallicity
(between -0.25 and 0.25) and potentially covering a wide range of ages.
A detailed comparison of the models with a NICMOS data set covering a 
much wider range of \vmini\ colors will be presented in 
Jensen et al. (2000).

\section{The F160W SBF Hubble Diagram\label{hubblesection}}

\subsection{Distances and Uncertainties}

To determine the distance to each galaxy, we adopted 
\Mbar\,${=}\,-4.86\pm0.05$ 
and computed the $k(z)$-corrected distance modulus:
\begin{equation}
(m{-}M) = \overline m_{\rm F160W} - \overline M_{\rm F160W}-k(z). 
\end{equation}
Redshift corrections $k(z)$ to fluctuation magnitudes 
in the F160W-band were taken from the Liu et al. (2000) models for 
metallicities between [Fe/H]\,${=}\,{-}0.4$ to 0.0 and old stellar 
populations.   
$k(z)$ corrections are listed in Table~\ref{distancetable}.
We compared these $k(z)$ corrections to those determined using Worthey's
models for solar metallicity (G. Worthey, private communication) 
and found small differences of order ${\lesssim}0.01$ mag.  
At distances of 10,000 \kms\ and less, the magnitude of the $k(z)$ corrections
are insensitive to the details of the stellar population models. 

\begin{deluxetable}{lr}
\tablecolumns{2}
\tablewidth{0pc}
\tablecaption{Typical Uncertainties \label{errorbudget}}
\tablehead{
\colhead{Source} &
\colhead{$\sigma$}
}
\startdata
\sidehead{\it Random Uncertainties}
PSF normalization and fitting\dotfill & 0.06 mag \\
Sky subtraction\dotfill   & 0.01 mag \\
Globular cluster and background galaxy removal\dotfill & 0.07 mag\\
Galaxy profile subtraction\dotfill & 0.02 mag \\
Bias subtraction\dotfill & 0.01 mag\\
Wormy background correction (see text)\dotfill & ${\sim}0.08$ mag\\
\Mbar\ calibration \dotfill& 0.05 mag\\
CMB Velocities\dotfill & 200 \kms \\
\sidehead{\it Systematic Uncertainties}
NICMOS photometric zero point\dotfill & 0.02 mag\\
Cepheid distance calibration\dotfill & 0.16 mag\\
\enddata
\end{deluxetable}

The uncertainties in the fluctuation magnitudes in
Table~\ref{distancetable} are the contributions from PSF fitting,
sky subtraction, bias removal, and galaxy 
subtraction, all added in quadrature.  
The uncertainties in the distance moduli include the uncertainty 
in \Mbar\ of 0.05 mag.
Typical values for the individual uncertainties are listed in 
Table~\ref{errorbudget}.  
We also determined the range of fluctuation
magnitudes permitted given the level of worminess in
the background and the agreement between individual annuli.  
Maximum and minimum values derived from the residual background
corrections are listed separately from the other sources of 
uncertainty in Table~\ref{distancetable}.

We treated the different uncertainties as if
they were independent, but acknowledge the fact that there are
subtle correlations between sources of uncertainty that
are difficult to quantify.  For example, the procedure that is
used to fit and subtract the smooth galaxy profile is affected
by errors in sky subtraction.  
While relationships between sources of uncertainty exist,
they are insignificant to the results of this study.  
Examination of Table~\ref{errorbudget} shows that the 
significant sources of uncertainty in the distance measurement
are the PSF fit, globular cluster correction, and
the intrinsic scatter in the \Mbar\ calibration.  
The first is due mainly to 
variations in the drift and focus of the telescope and brightness of
the PSF stars.  The second is principally a function of the depth of
the observation and size of the globular cluster population.
The cosmic scatter in \Mbar\ is a result of variations in the 
stellar populations of galaxies.  
These uncertainties are independent and may safely be added in
quadrature.
Furthermore, in many cases even these uncertainties are secondary to 
the larger uncertainty in the correction for worminess in the background,
which is a function of time since the last SAA passage.

Measuring the uncertainty due to residual background patterns was 
difficult; to make an estimate, we explored the range of 
correction that is permitted by the data by subtracting various levels 
of uniformly distributed residual spatial power, and thereby found the 
maximum and minimum fluctuation magnitudes allowed.  The most 
likely fluctuation magnitude for each galaxy was determined 
taking into account the details 
described in the notes in the appendix.  Rather than 
assume a Gaussian distribution of errors about the most probable value, 
we chose to adopt a probability distribution that increases linearly from 
zero at the maximum and minimum allowed values to the most likely 
value and is normalized appropriately.  The probability function is not 
symmetrical about the most likely value because the measurement is not 
usually midway between the maximum and minimum allowed values.  
We convolved this skewed saw-tooth distribution function with the 
normal probability distribution of the other sources of uncertainty
to get the probability distribution function that was used to
determine \Ho.

Some systematic errors listed in Table~\ref{errorbudget} 
affect all our measurements equally, and are
not included in the uncertainties in Table~\ref{distancetable}.
The first of these is the 0.02 mag uncertainty in the photometric
zero point of the F160W filter in the NIC2 camera.
The other systematic errors we inherit from the Cepheid distances
adopted, either directly or via the $I$-band SBF calibration.
The systematic uncertainty in the Cepheid distance scale of 0.16 mag
includes the 0.13 mag uncertainty in the distance to the LMC and the 
0.02 mag uncertainty in
the zero point of the period-luminosity relationship for Cepheid variables.  
The systematic photometric uncertainty in the WFPC-2 measurements
contributes another 0.09 mag to the Cepheid distances.
A detailed discussion
of these uncertainties can be found in Ferrarese et al. (2000a).
Adding all sources of systematic uncertainty in quadrature gives 0.16
mag.  An additional systematic uncertainty from the $I$-band SBF
distance scale is not included because the $I$-band distances are only
used to link the Cepheid calibration to the distant galaxies of our
sample.  The $I$-band SBF systematic uncertainties are the same
as those already discussed, and it would not be correct to include
them twice.  The 0.01 mag agreement between the $I$-band SBF and the
direct Cepheid calibrations confirms that no additional systematic error
is incurred by adopting the $I$-band SBF distances for the calibration.

\subsection{Radial Velocities}

The heliocentric velocity for each cluster or galaxy was initially measured
or collected from the literature by Postman \& Lauer 
(1995, and references therein).  
The uncertainties on the individual redshift measurements were 
typically 60 \kms.  
New data now available provide velocities to additional cluster members 
and have been included in this study.  
Radial velocities are compiled in Table \ref{distancetable}, along with
the number of individual galaxy redshifts that were averaged to get the
cluster velocity.  The details of how galaxies were selected for inclusion 
are described by Postman \& Lauer (1995).  
The mean uncertainty in the mean cluster
redshift is 184 \kms\ for the Postman \& Lauer sample.  NGC~4709 is listed 
in Table \ref{distancetable} with its own radial velocity; 
it is a member of the high-velocity
(4500 \kms) component of the Centaurus cluster, and hence has a significant
peculiar velocity. 
NGC~5193 is also listed with its own redshift; it was previously thought
to be the cD galaxy in Abell 3560, but  
Willmer et al. (1999) found that it is in fact a foreground galaxy.
NGC~4073 is not associated with a cluster; its heliocentric velocity
was taken from Beers et al. (1995).  
The heliocentric velocities were converted to the reference frame 
that is at rest with respect to the cosmic microwave
background (CMB) radiation.  The CMB dipole adopted was that measured
by Lineweaver et al. (1996).  

\subsection{The Model Velocity Field and \Ho \label{flowsection}}

Measurements of the Hubble constant within 50 Mpc must take
peculiar velocities into account because they can be a significant 
fraction of the Hubble velocity.  In fact, one of the differences
between the Hubble constants measured by Tonry et al. (SBF-II)
and Ferrarese et al. (2000a) using the {\it same} Cepheid calibrators 
and the {\it same} SBF measurements (albeit with a slightly different
calibration) was the result of different assumptions 
about the local velocity field.  We have chosen our distant sample
to be distributed in such a way as to minimize sensitivity to local
peculiar velocities (Fig.~\ref{supergal}).  
By far the greatest immunity to peculiar velocities
comes from reaching much greater distances than previously possible.
At 130 Mpc, we expect peculiar velocities to be approximately 3\%
of the Hubble velocity.  This insensitivity to peculiar velocities
and isotropic distribution of the distant sample produced the
most accurate SBF measurement of \Ho\ to date.

We followed the SBF-II maximum-likelihood procedure for computing \Ho. 
We first constructed a model velocity field, which 
included a 187~\kms\ cosmic thermal velocity dispersion.  
The quadrupole term adopted in SBF-II was not included.  
Various dipole terms (resulting from the peculiar velocity of the Local 
Group in the CMB frame) were tried, and a comparison is presented below.

At the position of each galaxy as defined by the F160W SBF distance,
the most likely velocity was determined from the velocity model.
A number of points were then chosen radially spanning the range of
possible distances given the uncertainties in the measured distances.
At each point, the joint likelihood of a given combination of
distance and velocity measurements was computed, and the likelihood
integrated across the radial range in distance.
The Hubble constant is a free parameter of the velocity model,
and this procedure was repeated to find the value of \Ho\ which
maximizes the likelihood of all the distance and velocity 
measurements together.
This procedure used the distance probability distribution function
constructed by convolving the normal Gaussian uncertainties with
the saw-tooth probability distribution between the maximum and 
minimum \mbar\ values.  
In practice, we attempted to {\it minimize} the negative 
likelihood statistic ${\cal N}$ (see SBF-II for details).
The value of $\chi^2$ 
determined using the maximum likelihood technique and our non-Gaussian
probability distributions is not necessarily minimized when 
${\cal N}$ is minimized; however, the difference
between values of ${\cal N}$ for different input parameters to the
velocity model is equivalent to a difference in $\chi^2$.
In Table~\ref{flowstable} we compared the likelihood of various
models by indicating the difference in $\chi^2$ relative to the
baseline model that ignores all peculiar velocities except the
motion of the Local Group in the CMB frame.

Several velocity models were used to determine the sensitivity of 
the \Ho\ measurement to the input parameters of the models.  
Results for these tests
are listed in Table~\ref{flowstable}.  NGC~4709 was excluded
from all fits because the velocity field of the complex Centaurus
cluster was not included in the velocity model. 
The models tried were constructed as follows:

\begin{deluxetable}{clclcrl}
\small
\tablecolumns{7}
\tablewidth{0pc}
\tablecaption{Values of \Ho\ for Different Velocity Models\label{flowstable}}
\tablehead{
\colhead{Model} &
\colhead{Fixed Model} &
\colhead{Dipole} &
\colhead{Dipole $v$} &
\colhead{\Ho} &
\colhead{$\Delta\chi^2$ \tablenotemark{b}} &
\colhead{Reference} \\
&\colhead{Components\tablenotemark{a}}&
\colhead{$(l,b)$} &
\colhead{(\kms)} &
\colhead{\tiny(\kmsmpc)} & &
}
\startdata
(1) & CMB only\dotfill           & \nodata  &\nodata        & $76.1{\pm}1.3$ & 0.0 & Lineweaver et al. 1996\\
(2) & Virgo, GA \& dipole\dotfill& (306,43) & $205{\pm}83$  & $77.1{\pm}1.6$ &$-$0.2 & SBF-II\\
(3) & Willick \& Batra dipole\dotfill& (274,67) & $243$         & $75.6{\pm}1.3$ & 0.2 & Willick \& Batra 2000\\
(4) & Giovanelli et al. dipole\dotfill& (295,28) & $151{\pm}120$ & $75.5{\pm}1.3$ & 0.4 & Giovanelli et al. 1998\\
(5) & Virgo, GA \& dipole\tablenotemark{c}\dotfill& (355,56) & $409{\pm}335$ & $76.9{\pm}1.5$ & 1.4 & SBF-II+free dipole \\
(6) & Lauer \& Postman dipole\dotfill& (343,52) & $689{\pm}178$ & $73.8{\pm}1.5$ & 9.7 & Lauer \& Postman 1994 \\
(7) & Model (2) + Bubble\tablenotemark{d}\dotfill& (306,43) & $205{\pm}83$  & $72.3{\pm}2.3$ & \nodata & Zehavi et al. 1998\\

\enddata
\tablenotetext{a}{All models include a cosmic velocity dispersion of 187 \kms\ as in SBF-II.}
\tablenotetext{b}{The relative likelihood of each model is quantified here as a difference
in units of $\chi^2$ from the baseline CMB model (1).}
\tablenotetext{c}{The dipole term is a free parameter in this model, and $\Delta\chi^2$
has been adjusted to account for the three additional degrees of freedom. 
The positions of the Virgo and Great Attractor clusters are fixed as in SBF-II, without the quadrupole term.}
\tablenotetext{d}{Model components are the same as in (2), but only the six 
galaxies more distant than 96 Mpc outside the putative locally under-dense region are used to 
compute \Ho.}
\end{deluxetable}

(1) The first model does not include any local attractors or peculiar
velocities beyond that of the Local Group in the CMB frame.  
In the CMB frame, \Ho${=}76.1$ \kmsmpc.  We adopt the
CMB model as the baseline and compare other models by computing
the change in $\chi^2$ relative to this case.

(2) Adding the contributions from the Virgo and GA mass concentrations
and dipole as prescribed by SBF-II increases \Ho\ to 77.1.
The slight decrease in $\chi^2$ is not significant.
The quadrupole term suggested by SBF-II was not included because
it is inappropriate for the distances of the galaxies in our sample
(including it would increase $\chi^2$ by 25!).

(3$-$4) For model 3 we used the dipole determined by 
Willick \& Batra (2000). 
Model 4 includes the dipole measured by Giovanelli et al. (1998) using 
Tully-Fisher measurements to many clusters out to redshifts of 9000 \kms.  
The likelihood of these two dipole models is essentially the same as the
best-fitting SBF-II models and the CMB-only baseline model.
The Hubble constant implied by these models is approximately 75.5 \kmsmpc.
The largest difference in \Ho\ between models 1 to 4, which have 
essentially the same likelihood, is only 1.6 \kmsmpc.  

(5) Like model 2, the fifth model used the mass distribution 
suggested by SBF-II, 
but allowed the maximum likelihood procedure to 
determine the most likely dipole velocity in addition to \Ho.  
Despite having more freedom to fit the data with three additional
degrees of freedom, the fit is worse and and $\Delta\chi^2$ is larger.
The dipole determined is large, but barely larger than the 
uncertainty.  We have sampled the velocity field with
only 16 points spanning a range in distance from 50 to 150 Mpc, and
the sample was chosen to minimize sensitivity to streaming motions
that could bias the measurement of \Ho.  Towards this end we were
successful; the small variation in \Ho\ between the various models
confirms this conclusion.  On the other hand, to reliably measure
the bulk motion of the galaxies in the local universe, 
distances would need to be
measured to a much larger sample of galaxies within the redshift
interval of interest (7000 to 10,000 \kms\ in this study).  

(6) The one dipole presented here that fails to
fit our data very well is the large dipole velocity of 689 \kms\
measured by Lauer \& Postman (1994).  Using the Lauer \& Postman 
dipole reduces \Ho\ by approximately 3 \kmsmpc, but $\chi^2$ 
is significantly larger.  It is, however, closer to the free dipole 
of model 5 than the other model dipoles considered.

\begin{figure}
\epsscale{0.8}
\plotone{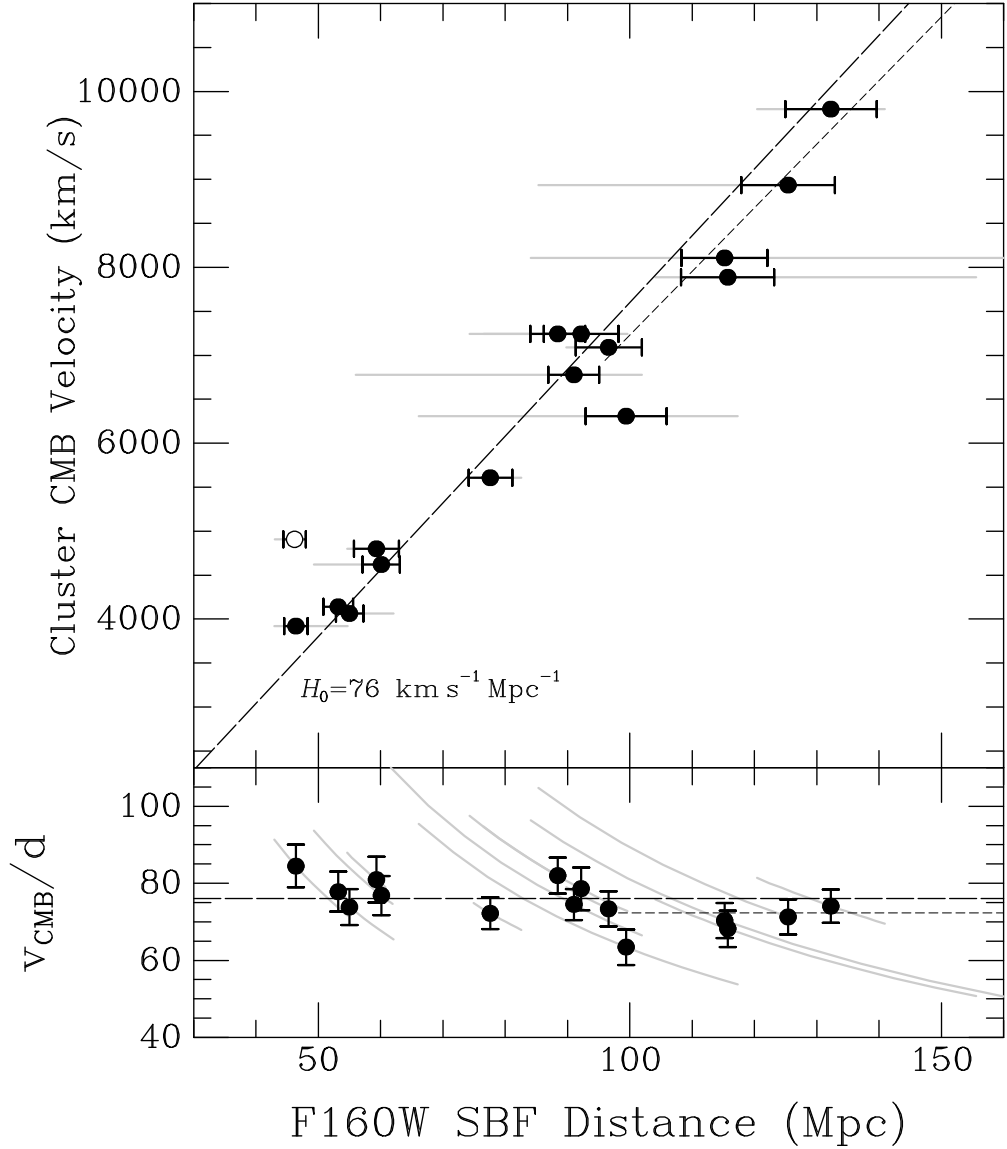}
\figcaption[]{The Hubble diagram is plotted for the sample galaxies
using their velocities in the CMB frame.
The 1-$\sigma$\ error bars shown indicate the 
contributions from the scatter in the F160W SBF calibration, the PSF 
normalization, sky subtraction, galaxy subtraction,
bias subtraction, and globular cluster removal, all added in quadrature.  
The underlying gray lines indicate the full
range of allowed distances as quanitified in Table \ref{distancetable}
as ``max'' and ``min'' values.
Our best-fit Hubble constant of 76 \kmsmpc\ is indicated; 
NGC 4709 (open symbol) is excluded from the fit.  The second
dashed line at \Ho$\,{=}\,72.3$ \kmsmpc, determined using only the
six galaxies more distant than 96 Mpc, 
shows the 6\% reduction in \Ho\ beyond 
70~h$^{-1}$ Mpc suggested by Zehavi et al. (1998).
The lower panel shows the Hubble ratio $v/d$ for each point (including
a nominal 200 km/s uncertainty in the velocities for illustrative
purposes only).  The allowed ranges are 
plotted as dotted curves, indicating how $v/d$ changes with a change
in the adopted distance for a particular galaxy. 
The lower value of \Ho\ for the outer six points is also 
shown.
\label{hubblediagram}}
\end{figure}

Based on the full data set, we conclude that
\begin{equation}
H_{0} = 76 \pm 1.3\ ({\rm random}) \pm 6\ ({\rm systematic})\
{\rm km\,s}^{-1}\,{\rm Mpc}^{-1}.
\end{equation}
The 1-$\sigma$ random uncertainty formally includes all sources of 
uncertainty in the distance measurement, including the non-Gaussian 
uncertainty from the residual background correction.  
Although $\chi^2$ per degree of freedom is not minimized for our
non-Gaussian probability distribution, its value of 1.0 for model 1
indicates that the adopted uncertainties are reasonable. 
The Gaussian 1-$\sigma$
error bars are plotted with thick lines in the Hubble diagram
(Fig.~\ref{hubblediagram}).  The full non-Gaussian ranges 
allowed by the various corrections to the SBF distances are indicated
by the lighter lines underneath each point.   The Hubble diagram
is shown using the CMB velocities (model 1), and our best-fit value 
of \Ho${=}76$ \kmsmpc\ is indicated by the dashed line.  

A second line in
Figure~\ref{hubblediagram} indicates the decrease in \Ho\
beyond 70 h$^{-1}$ Mpc (${\sim}7000$ \kms) suggested by Zehavi et al. (1998).
Their ``Hubble bubble'' model hypothesizes that a locally under-dense
region of the Universe gives rise to 
an expansion rate about 6\% higher within 
70 h$^{-1}$ Mpc.  The SBF analysis was repeated using only the
six most distant galaxies and the model 3 (SBF-II) velocity 
model (model~7 in Table~\ref{flowstable}).  The six galaxies
were chosen to minimize \Ho\ and provide the best match to the 
decrease in \Ho\ predicted by Zehavi et al. (1998).
The result, 72.3 \kmsmpc, reproduces nearly perfectly the
predicted decrease in \Ho.

The lower panel in Figure~\ref{hubblediagram} shows the Hubble ratio
$v_{\rm CMB}/d$ for each galaxy.  The error bars are shown for the
Gaussian component of the distance error, and do not include the 
uncertainty range from the residual background correction.  
The curved lines behind each point show the full range of 
possible distances if a different background correction $P_g$ were
adopted.  The longest arcs are necessarily those points with the
worst worminess, the largest corrections, and the lowest S/N ratios.
The best-fit value of \Ho$\,{=}\,76$ \kmsmpc\ is indicated by the 
horizontal line.  Once again, the Zehavi et al. (1998) predicted
decrease in \Ho\ is shown.

We also explored the sensitivity of our results to the low-S/N observations. 
Several of the measurements are quite poor, and should arguably be
excluded from the fits.  Excluding all galaxies obviously
contaminated by worms (Abell 262, 2666, 3389, 3581, 3742 and NGC 4073) gives
\Ho$\,{=}\,77.4\pm1.7$ \kmsmpc\ for model 2, which is nearly the same
value determined using the entire data set of 77.1. 
If we exclude only the three galaxies with
$\xi{<}1$ (Abell 2666, Abell 3389, and Abell 3581), then we find
that \Ho$\,{=}\,78.0\pm1.6$.  It is clear that the worst measurements
are not systematically biasing the measurement of \Ho.  This is
not surprising, as the maximum likelihood code takes into account
the large range of possible distances for these galaxies.  The fact
that there is no systematic offset shows that the $P_g$ corrections
are applied uniformly and that the residual background is not systematically
over or under-subtracted.
When the six galaxies that have dust lanes or disks are exluded 
(Abell 262, 1060, 2199, 3565, 3581, and NGC 5193), 
\Ho$\,{=}\,75.7\pm1.5$ \kmsmpc.
This value is only 1.4 \kmsmpc\ smaller than 77.1, and suggests that clumpy
dust in some galaxies does not introduce a significant bias to
the distance measurements.
If there were a significant bias in the distance measurements due to 
residual cosmic rays, clumpy dust, improper bias removal, undetected
globular clusters, or any of the other sources of variance discussed 
above, the result would be increasingly underestimated distances as
redshifts increase.  In fact, the opposite trend is observed:  the 
highest redshift galaxies have somewhat larger measured distances
than expected, as predicted by the Zehavi et al. (1998) Hubble bubble
model.

\section{Can \Ho\ be 65 \kmsmpc?\label{whatifsection}}

Several groups have recently reported measurements of the Hubble constant
derived from HST Cepheid distance calibrations of various secondary
distance indicators.  Our best-fit measurement of 
\Ho$\,{=}\,76\,\pm\,1.3\,\pm\,6$
\kmsmpc\ is in good agreement (better than 1\,$\sigma$) with several,
including optical SBFs (SBF-II; Lauer et al. 1998), Cepheid
distances alone (Willick \& Batra 2000), fundamental plane distances
(Kelson et al. 2000), and Tully Fisher distances (Sakai et al. 2000).
The SBF Hubble constant as calibrated by Ferrarese et al. (2000a) differs
from ours at the 1.5-$\sigma$ level for the same reasons it differs
from SBF-II:  a slightly different SBF calibration and different peculiar 
velocities were adopted for the four clusters measured by Lauer et al. 
(see Ferrarese et al. for a discussion).
Our measurement of \Ho\ differs significantly from
the results based on type-Ia supernovae.  Gibson et al. (2000) report
\Ho$\,{=}\,68\pm2\pm5$ \kmsmpc, 12\% lower than our value.  Parodi et al.
(2000) found \Ho$\,{=}\,58.5\pm6.3$ \kmsmpc\ (90\% confidence level), 
which is 24\% smaller. These two supernovae measurements cannot be directly
compared, however, because of a significant difference
(0.18 mag) between the calibration adopted by Parodi et al.
and that used by the HST Key Project team (ie., Gibson et al.)
The Key Project Cepheid calibration was adopted for this study.

Is it possible that \Ho$\,{=}\,65$ \kmsmpc, and that we have 
overestimated it by 15\% (or more)?  Roughly one third of this 
difference disappears if the Zehavi et al. (1998) Hubble bubble
model is correct.  If galaxies nearer than ${\sim}7000$ \kms\ 
must be disregarded because of their enhanced velocities away from a 
locally under-dense region, then our measurement of \Ho\ can be as 
low as 72 \kmsmpc.  
Our results do not demand that this be the case, however, and are
still consistent with observations that refute the Hubble
bubble hypothesis (Giovanelli et al. 1999, Lahav 2000).  
The fits including all the data, assuming a smooth Hubble flow,
are equally good because the nearer meausurements tend to be most reliable.  

Extinction from clumpy dust in the individual galaxies could add to
the fluctuation power and bias the SBF measurements to shorter
distances.  
No sign of dust was seen in the images of the galaxies used to
determine the calibration of \Mbar\ (aside from NGC~1387 and NGC~4536,
which were excluded).  Six of the distant galaxies
do have obvious dust lanes (Table~\ref{powertable}).  
Only one (Abell 262) has extensive dust, and we used the optical 
WFPC-2 image of Lauer et al. (1998) to mask the dusty regions.
The other five have dust lanes or disks that are concentrated near the 
centers of the galaxies, which were masked. 
There is no evidence for extended patches of dust with sizes comparable
to the PSF and smooth on larger scales.  
The distances for these are not systematically smaller than the others 
at the same redshifts, nor are their colors redder on average.
The Hubble constant measured with the six dusty galaxies excluded was not
significantly smaller, and  
it seems unlikely that all the distant galaxies would have uniformly 
distributed clumpy dust that would be unrecognizable in our images.

Besides the potential 6\% reduction in \Ho\ beyond ${\sim}100$ Mpc
suggested by the six most distant measurements,
are there systematic problems with the F160W SBF measurements 
that could explain another ${\sim}$10\% (or more) difference 
between our results and the conclusions of the supernovae measurements?
There are potentially three sources of extra power in the power 
spectrum that are not convolved with the diffraction pattern of the 
telescope, but do have power on the spatial scales over which we fit the 
SBF power spectrum.  
The first of these is the residual bias scaled by the flat field image.
We addressed this possibility by explicitly subtracting a 
dithered bias$\times$flat image as described in 
Section~\ref{reduxsection}.  
The uncertainty in \Ho\ resulting from errors in bias subtraction 
was measured and found to be less than 1\%
(Table~\ref{errorbudget}).

The second potential contributor to the power spectrum is the residual
wormy background. 
We carefully excluded wormy images and subtracted an estimate 
of the residual power as described in Section~\ref{reduxsection}.  
If the correction for worms were 
systematically underestimated, then our measurement of \Ho\ would be too 
large.  In the previous section we showed that excluding the galaxies
contaminated by worms had no significant effect on the measurement
of \Ho.  Excluding the lowest S/N measurements also had no 
significant effect.  The range of \Ho\ values seen during these 
tests was less than 1 \kmsmpc.
This suggests that the corrections were 
applied uniformly.

The third potential contributor to the power spectrum is the 
residual structure in the background from subtraction of the 
model galaxy profile.  
To avoid any bias because of the somewhat arbitrary fit of the 
left-over large-scale structure in the galaxy, we excluded
wavenumbers smaller than 20 from our analysis.  
The mean uncertainty in the distance modulus from galaxy and smooth 
background subtraction was 0.02 mag.  Residual galactic structure
could explain a 1\% bias in our \Ho\ measurement, but not a large
systematic error.

If the fluctuation powers we measure are reliable, is it possible that other 
sources of systematic error could cause us to overestimate \Ho\ 
by 10 to 15\%?
Perhaps the most obvious candidate for this kind of systematic error 
would be the calibration of the F160W absolute fluctuation magnitude.  
We used both Cepheid and $I$-band SBF distances to 
determine the calibration.  Although not completely independent, the 
two calibrations are remarkably consistent (0.01 mag).  
The agreement between the $I$-band SBF and direct Cepheid calibration 
supports the conclusion that there is no significant difference in the 
fluctuation amplitudes between early and late-type galaxies.  
The applicability
of the calibration to the more distant galaxies is demonstrated by the
overlap with the intermediate-distance sample.  
To explain a 15\% difference in \Ho,
\Mbar\ would have to be brighter by 0.3 mag, or \Mbar$\,{=}-5.16$.  
Examination
of Figure~\ref{calibration} shows that a calibration as bright as
$-$5.16 is inconsistent with the data.  

Significant systematic errors in the Cepheid distance scale are relevant 
to the measurement of the true value of the Hubble constant, but cannot
explain the difference between our measurement and that determined
using type-Ia supernovae because we adopted the same Cepheid calibration as
the other groups listed at the beginning of this section.  
A systematic error in the distance to the LMC (for example) will affect 
our measurement of \Ho\ in exactly the same way as the other measurements.

Is it possible that the mundane choice of Galactic extinction corrections 
could result in a systematic calibration error at the 15\% level?  
By observing in the near-IR, our sensitivity to errors in the 
extinction are significantly reduced.  The largest correction in our 
sample is 0.08 mag (Table~\ref{sampletable}).  Most of the calibrator
galaxies have IR extinction corrections of order 0.01 mag.  
If extinction has been underestimated, the true fluctuation magnitudes
will be brighter than we have estimated and the distances smaller.
Increasing extinction corrections makes \Ho\ larger.  
On the other hand, extinction cannot have been overestimated by very much
because the corrections are already very close to zero. 
The only other way to get distance measurements that are systematically 
underestimated by 15\% would be for extinction corrections to all three 
Cepheid calibrators used in this paper and all the Cepheid calibrators 
used by Tonry et al.
(SBF-II) to be overestimated by 0.3 mag, but not those used to
calibrate the supernova distance scale.  It seems unlikely
that Galactic extinction could be the cause of so large a systematic error.

One reason that previous measurements of \Ho\ have disagreed with 
each other has been the choice of velocities used 
(Ferrarese et al. 2000a, Mould et al. 2000).  When \Ho\ is 
measured on scales where the peculiar motions of individual galaxies
are a significant fraction of the Hubble velocity,  
the value of \Ho\ will depend quite sensitively 
on the velocity adjustments made for infall into local mass concentrations.
Our measurement of \Ho\ reaches well into the Hubble flow and is
distributed uniformly on the sky, and is therefore very insensitive 
to the choice of velocity model and the peculiar velocities of individual
galaxies and clusters (as described in Section~\ref{flowsection}).

Finally, we cannot rule out the possibility that modest systematic 
errors affect both F160W SBF and type-Ia supernovae distance 
measurement techniques in such a way to create the difference between 
the measurements.

\section{Summary}

We measured accurate IR SBF distances to a collection of
16 uniformly-distributed distant galaxies for the purpose of measuring
the Hubble constant well beyond the influence of local peculiar velocities.
These NICMOS measurements mark the first time SBFs have been measured in 
galaxies out to redshifts of 10,000 \kms, clearly demonstrating the 
advantages of measuring SBFs in the near-IR with excellent spatial
resolution and low background.  
The calibration of the F160W SBF distance scale presented here was based
on SBF measurements of galaxies in which Cepheid variable stars were 
detected in the same galaxy.
Using a maximum-likelihood technique to account both for the influence
of local mass concentrations on the velocity field and the 
non-Gaussian uncertainties on our SBF distance measurements yields a 
Hubble contsant of \Ho$\,{=}\,76\pm1.3$
\kmsmpc\ (1-$\sigma$ statistical uncertainty) with an additional systematic
uncertainty of 6 \kmsmpc, primarily the result of uncertainty in the
distance to the LMC. 
The small statistical uncertainty in \Ho\ is a result of 
the fact that our measurement is very insensitive to peculiar velocities,
stellar population variations, extinction corrections, and photometric
errors.
Arbitrarily excluding all but the six most distant
galaxies from the fit results in a 6\% decrease in \Ho, 
consistent with the hypothesis
that the Local Group is located in an under-dense region of the universe.

\acknowledgements
This study benefitted greatly from NICMOS data collected as part of several 
programs, and we thank those who worked to acquire that data.  
In particular, we are grateful to those who helped ensure that the
data would be appropriate for SBF analysis and assisted with the
data reductions (D. Geisler, J. Elias, J. R. Graham, and S. Charlot).
We are endebted to the NICMOS GTO team for their hard work in building,
calibrating, and providing software for NICMOS.  
We greatly appreciated the helpful comments of R. Weymann, J. Blakeslee, 
and L. Ferrarese.
The calibration presented 
here made use of data collected by the Optical SBF
team (J. Tonry, J. Blakeslee, E. Ajhar, and A. Dressler), and we thank
them for providing color photometry and $I$-band SBF distances to the 
calibration galaxies.  
Finally, we wish to thank G. Worthey, S. Charlot, and A. Vazdekis for 
constructing stellar population models and providing appropriate 
$k(z)$-corrections.  

This research was supported in part by
NASA grant GO-07453.0196A.  The NICMOS GTO team was supported by
NASA grant NAG 5-3042.
J. Jensen acknowledges the support of the Gemini Observatory,
which is operated by the Association for Research in Astronomy, Inc.,
under a cooperative agreement with the National Science Foundation on
behalf of the Gemini partnership: the National Science Foundation
(United States), the Particle Physics and Astronomy Research Council
(United Kingdom), the National Research Council (Canada),  CONICYT
(Chile), the Australian Research Council (Australia), CNPq (Brazil)
and CONICET (Argentina).

\appendix
\section{Appendix:  Notes}

Abell 262 (NGC 708):  The central galaxy in Abell 262 is littered with 
dust.  We used the high-resolution $I$-band WFPC-2 images (Lauer et al. 
1998) to identify dusty regions and create a mask for our NICMOS 
image.  In addition to 
the copious dust, we had to exclude exposures because of 
worminess in the background.  The uncertainty in fluctuation 
magnitude is relatively large because of the dustiness and wormy background 
corrections, even though Abell 262 is among the closest of the clusters 
we observed.

Abell 496 (PGC 015524):  This cluster is the most distant in our 
sample, and we allocated 3 orbits to ensure a good SBF measurement.  
Of the 20 individual exposures, only the last two were found to be wormy.  
The other 18 are unaffected.  The S/N is good and the fluctuation 
measurement is reliable.

Abell 779 (NGC 2832):  Aside from a little dither-pattern noise 
in the power spectrum, the results for Abell 779 are quite good.  
Pattern noise is an array of spots in the spatial power spectrum with
a periodicity corresponding to the 20-pixel offset of the dither
pattern.  Detector artifacts (e.g.,  vertical bands or mismatches
in the background level at quadrant boundaries) were sometimes 
incompletely removed by the image reduction procedures and cause
pattern noise.  Pattern
noise is only significant in the power spectrum of the outermost annulus.

Abell 1060 (NGC 3311):  The central galaxy in the Hydra cluster was
one of four galaxies presented here that were observed by D. Geisler, 
J. Elias and E. Ajhar as part of NICMOS program 7820.
The images were reduced for SBF analysis using the software and procedures
described in this paper.
NGC 3311 has some dust in the central region that was masked; the SBF
fit in the outer regions is nearly perfect and the S/N ratio is 
very high.  

Abell 1656(a) (IC 4051):  Two galaxies in the Coma cluster were
observed by D. Geisler et al.  
IC 4051 has an unusually large population of 
globular clusters (Baum et al. 1997).  We found that many are much 
brighter than expected for a galaxy at this distance.
We modified the luminosity fitting parameters and estimated the 
contribution from unresolved GCs and subtracted it, but a relatively 
large uncertainty in the GC contribution to the SBF power remains.  
The fit to the SBF power spectrum is good.

Abell 1656(b) (NGC 4874):  
The power spectrum for NGC 4874 is clean and the fit is very good.  
NGC 4874 has a normal globular cluster population (Harris et al. 2000).

Abell 2199 (NGC 6166):  The central galaxy in Abell 2199 has 
dust lanes within 3 arcsec of the center.  
The dust lanes were masked prior to performing 
the SBF analysis, but measurements in the innermost aperture are 
suspect.  The SBF analysis did not include the region between 
NGC 6166 and two nearby companions, where the fit to the galaxy profile 
is not very good.  The fit to the power spectum in the intermediate
annulus was excellent, but the outer two apertures disagree at 
a level (0.64 mag) that cannot be corrected properly by adopting
a value of $P_g$ that scales with area.  We adopted the measurement
in the intermediate aperture and a relatively large range of permitted
fluctuation magnitudes.

Abell 2666  (NGC 7768):  NGC 7768 has surprisingly few globular 
clusters, which is consistent with the measurements of 
Harris, Pritchet, \& McClure (1995) and 
Blakeslee, Tonry, \& Metzger (1997).
Our attempts to fit a luminosity function to a half-dozen 
objects failed to produce a reasonable correction for undetected 
globular clusters.  The SBF analysis proceeded without a GC 
correction, and we adopted an uncertainty that is larger than the other 
galaxies that reflects our lack of knowledge of the GC luminosity 
function.  The only way this measurement could be significantly biased 
by undetected GCs is if the GC luminosity function is skewed to the 
faint end and contains practically no GCs on the bright side of the peak.  
The Abell 2666 observation is 
also contaminated by a wormy background.  One of the six exposures is 
excluded, and the worst regions masked in the two subsequent 
exposures.  There is always a tradeoff between including 
frames that increase the SBF signal but also contain the decaying 
wormy background.  In this case, a good fit to the power spectrum was 
achieved, but a significant correction for the background must be 
applied to make the outer two apertures agree.  $P_g$ is further
enhanced by the presence of undetected globular clusters and background
galaxies that could not be handled with the usual procedure of fitting
luminosity functions due to the paucity of bright objects in this 
field.  $\xi{<}1$ for this galaxy and the possible range of 
fluctuation magnitudes is therefore quite large.

Abell 3389 (NGC 2235):  We observed the central galaxy in Abell 
3389 in the continuous viewing zone to achieve a longer total 
integration time for this galaxy.  Unfortunately, the longer MULTIACCUM
sequences used to avoid frequent NICMOS buffer dumps had many more 
persistent cosmic rays in each image and a significantly wormy background.  We 
abandoned half of our images, and the remaining ones must be 
corrected for residual worminess.  As a result, $P_g$ is significant and
$\xi$ is less than unity. 
The fluctuation power increases significantly with aperture area, and the 
globular cluster and worminess corrections are large.  The uncertainties 
reflect the fact that the fluctuation magnitude is poorly constrained.

Abell 3565 (IC 4296):  IC 4296 has a compact dust ring close to the 
nucleus, but no sign of dust outside of a radius of 1.5 arcsec.  
Very small residual spatial variance corrections ($P_g$) 
bring the annuli into nearly perfect agreement.
The S/N of this measurement is high and the GC correction small.

Abell 3581 (IC 4374):  the observations of IC 4374 were strongly 
affected by worminess in the background.  Two of six exposures were 
excluded from the final image, and residual worms were masked in 
three of the remaining four.  The potential for bias is strong in this case, 
and a significant $P_g$ correction for background power was applied,
resulting in a $\xi{<}1$ and a large range of allowed fluctuation
magnitudes.  
Furthermore, the central regions contain a dust lane, which we masked. 

Abell 3656 (IC 4931):  The only problem that arose in the analysis of 
IC 4931 is the presence of dither pattern noise in the background.  
This problem is only significant in the largest 
annulus; the inner two agree nicely.  

Abell 3742 (NGC 7014):  The HST failed to lock onto the guide stars 
for the observations of NGC 7014.  Because some drift occurred during 
the MULTIACCUM sequences, our IDL procedures interpreted the 
changing flux in each pixel as cosmic rays.  To overcome this problem, 
we were forced to abandon the temporal cosmic ray rejection and rely 
on the spatial information alone.  The galaxy fitting routine also had 
trouble because of the smeared image.  The residual image shows extra 
background structure close to the center where the galaxy fit is worst.  
In this case, we included a small
correction to the fluctuation power that scales as the galaxy brightness 
(rather than by the area of the aperture, as with the worminess 
corrections applied to some of the other galaxies).  Furthermore, the 
S/N is reduced because the PSF has been smeared by telescope drift.  
Because the bright PSF stars used for the other galaxies do not match in 
this case, we resorted to extracting a low-S/N PSF from a combination 
of six faint stars or globular clusters from the smeared image of NGC 
7014.  The resulting fit is acceptable, but the PSF normalization 
somewhat uncertain.  Although Abell 3742 is among the closest 
clusters in our sample, the uncertainties are relatively large.

NGC 4073:  This galaxy was observed by D. Geisler et al., 
and it is not associated with a cluster.  
Its globular cluster population is extensive.
Worminess in some of the images contaminates the SBF measurement,
and the $P_g$ correction is large. 
Another difficulty with the analysis of this galaxy was accurately 
subtracting the smooth galaxy profile.  
Because of the dither pattern used in this
case was chosen to maximize the number of globular clusters detected, 
there is a hole in the image near the galaxy center that made galaxy
subtraction somewhat difficult.

NGC 4709:  This galaxy is part of the complex Centaurus cluster and 
has a significant positive peculiar radial velocity.  
Although it is not useful for 
measuring the Hubble Constant, it does have a reliable $I$-band SBF 
distance from WFPC-2 observations.  The first of the six 
exposures was excluded because of low-level worminess in the 
background.  The resulting power spectrum fits the PSF spectrum very 
well, and the S/N is relatively high.  A modest correction for background 
worminess brought the individual apertures into perfect agreement 

NGC 5193:  A recent velocity study by Willmer et al. (1999) 
indicated that NGC 5193 is not a member of the 
Abell 3560 cluster, as had been supposed.   
NGC~5193 has a dust ring extending 2.6 arcsec from the center, which we 
masked.  NGC 5193 is one of the nearest galaxies in our 
survey, and the S/N ratio is quite high.  
Nevertheless, the relatively large range of acceptable fluctuation
magnitudes reflects the 0.5 mag disagreement 
between annuli that was not removed with a uniform $P_g$ correction.

\end{document}